\documentclass[aps,twocolumn,prc,showpacs,unsortedaddress,superscriptaddress]{revtex4}
\usepackage{url}
\usepackage{ulem}
\usepackage[usenames,dvipsnames]{color}
\usepackage{epsfig}
\usepackage{subfigure}
\usepackage{amsmath}
\usepackage{amssymb}
\usepackage{amstext}
\usepackage{slashed}
\usepackage{wasysym}
\usepackage{amsfonts}
\usepackage{mathbbol}
\usepackage{textcomp}
\usepackage{graphicx}
\usepackage{placeins}
\usepackage{dashrule}

\begin{document}

\title{Effect of the symmetry energy and hyperon interaction on neutron stars}

\author{Daniel Bizarro}
\email{danielbizarro@gmail.com}
\affiliation{Centro de F\'{\i}sica Computacional, Departamento de F\'{\i}sica,
Universidade de Coimbra, P-3004-516 Coimbra, Portugal}

\author{Aziz Rabhi}
\email{rabhi@teor.fis.uc.pt}
\affiliation{Centro de F\'{\i}sica Computacional, Departamento de F\'{\i}sica,
Universidade de Coimbra, P-3004-516 Coimbra, Portugal}\affiliation{University of Tunis El-Manar, Unit\'e de Recherche de Physique Nucl\'eaire et des Hautes \'Energies, Facult\'e des Sciences de Tunis, 2092 Tunis, Tunisia}

\author{Constan\c ca Provid\^encia}
\email{cp@teor.fis.uc.pt}
\affiliation{Centro de F\'{\i}sica Computacional, Departamento de F\'{\i}sica,
Universidade de Coimbra, P-3004-516 Coimbra, Portugal}

\begin{abstract}
The joint effect of the density dependence of the symmetry energy and
strangeness content on the structure of cold neutron
stars is studied within the framework of a relativistic mean field
theory. It is shown that 2$M_\odot$ are obtained for repulsive $YY$
interaction and preferably for a small or a large slope $L$. An attractive
$\Sigma$ potential in nuclear matter will favor the appearance of
strangeness in stars with a mass as small as $\sim 1 M_\odot$, if,
however it is repulsive only stars with a mass $\gtrsim 1.4 M_\odot$
will contain strangeness. 
The joint effect of reducing the symmetry
energy slope and including hyperons is to farther reduce the radius.
Neutron star maximum mass evolve non-monotonically with the symmetry
energy slope, and the smallest masses are obtained for values $L\sim
80$  MeV. Other neutron star variables evolve nonlinearly with the slope of the symmetry energy
and depend on the hyperon-nucleon and hyperon-hyperon couplings.
The radius of a neutron star is linearly correlated with the neutron star total strangeness fraction and
the slope is independent of the slope of the symmetry energy and the
mass of the star. 
\end{abstract}
\maketitle

\section{INTRODUCTION}
%\label{introduction}

The structure of neutron stars depends strongly on the equation of state (EOS)
of nuclear matter at suprasaturation densities \cite{lattimer_prakash_2001}.
The densities in the center of neutron stars can go up to 8 times the
nuclear saturation density. Consequently, the
nucleon Fermi energy rises above the rest mass of hyperons thus making their appearance in the
inner layers of the star {possible}. Hyperons were first
taken into account in the description of stellar matter in
\cite{ambart1960}.
Within a relativistic mean field (RMF) approach, hyperons have been first
included in the EOS of stellar matter   in  \cite{knorren_hyperons_1985,schaffner1996,gm1,glendenning_hyperons_1985}.
In
\cite{schaffner1996} besides the usual non-strange
mesons $\sigma,\omega$ and $\rho$,  the
vector meson $\phi$ and the scalar  meson $\sigma^*$  with hidden
strangeness were also included.

The recent observations of the high mass neutron stars PSR J1614-2230 \cite{197_neutron_star_demorest}
with $1.97\pm 0.04\,M_\odot$ and PSR J0348+0432 \cite{antoniadis2013}
with $2.01\pm 0.04\,M_\odot$ raises the question whether the interior
of compact star contains exotic degrees of freedom, in particular,
hyperons, kaon condensates or quark matter.
Calculations using the microscopic non-relativistic Brueckner-Hartree-Fock (BHF) formalism
taking into account free and interacting hyperons show that hyperons greatly soften the EOS,
significantly reducing neutron star masses, barely reaching the `canonical'
$1.4-1.5 M_\odot$   neutron star mass\cite{bhf}.
The authors of \cite{isaac11}, making use of reasonable assumptions, and
complementing the  BHF formalism with a density dependent Skyrme-like
term mimicking many-body interactions \cite{balberg_gal_effective_eos_1997}
to test the effect of three-body hyperonic forces,
were only able to reproduce up to $1.6\,M_\odot$ strange stars. This
mass will go up to 1.7 $M_\odot$ if a repulsive interaction is taken
for the $\Sigma$ hyperon \cite{domenicoPhD}.
However, in recent calculations using an auxiliary field  diffusion  Monte Carlo calculation \cite{lonardoni2013}, the authors have found that a strong repulsive three-body force, is needed to realistically describe the separation energy of the $\Lambda$-hyperon from hypernuclei, within a non-relativistic Hamiltonian.  This  repulsive $\Lambda$NN force produces a stiff enough equation of state of hyperneutron matter  that satisfies the 2 $M_\odot$ constraint,
shifting the onset of hyperons to densities above $0.56$ fm$^{-3}$ \cite{lonardoni2014}. 
Other studies were developed within a relativistic mean-field (RMF)
approach to discuss whether hyperons or other degrees of freedom  are
possible inside neutron stars
\cite{haensel11,schaffner12a,schaffner12b,sedrakian12,panda12,providencia13,colucci2013,debora14,raduta14}.
Masses of $\approx 2M_\odot$ were only achieved by
including the hidden-strangeness vector-meson $\phi$.

% symmetry energy  
The symmetry energy is relatively  well constrained  at nuclear
saturation density \cite{tsang12}.  However, an accurate characterization
of this property at all densities is needed
to properly describe asymmetric nuclear matter and stellar matter.
Neutron stars are very neutron rich so
we can expect that the symmetry energy plays
a significant role in this context. Many efforts have been made to understand the
density dependence of the symmetry energy ($\epsilon_\textrm{sym}$), see for
instance \cite{slope,Vinas2009,vidana09,camille11}.
Nuclear models are fitted to nuclear properties and this imposes 
correlations between several nuclear properties at saturation density,
such as the symmetry energy at saturation
$J=\epsilon_\textrm{sym}(\rho_0)$,  the slope of the symmetry energy
at saturation $L$,  the
curvature of the symmetry energy $K_\textrm{sym}$, 
and $K_\tau$ a term which characterizes the isospin dependence
of the incompressibility at saturation and subsaturation densities,
\cite{Vinas2009,vidana09,symmetry_energy_Fattoyev_2014}.
In the present work we investigate how the interplay between the
density dependence of the symmetry energy and the hyperon-nucleon and
hyperon-hyperon interactions define the structure and strangeness
content of a neutron star. We will work in the framework of a relativistic mean
field approach. The density dependence of the symmetry energy will be
modelized {by} introducing in the lagrangian density nonlinear terms that
include the vector isovector meson $\rho$, in particular, we will
include $\rho\sigma$ and $\rho\omega$ non-linear terms
\cite{horowitz2001a,horowitz2001b,carriere2003} and will allow
the slope of the symmetry energy at saturation to change from 50 to
110 MeV. This procedure
automatically establishes relations within the model between the different isovector
properties that will be discussed.

%comparing relativistic and non-relativistic approaches, with distinct experimental setups.

% our work
The
influence of density dependence of the symmetry energy   on the strangeness,
central density, gravitational mass and radius
of neutron stars will then be studied. 
Due to the lack of information on the hyperon-hyperon and hyperon-nucleon
interaction we will  consider a set of hyperonic
parametrizations which will take into account the existing uncertainties.
We will consider parametrizations which take into account YY
interaction through the inclusion of mesons with hidden strangeness
$\sigma^*$ and $\phi$ \cite{}. This will allow to include
extra repulsion at large densities if the coupling of the $\sigma^*$
meson to hyperons is not too strong.
%The modellization of
%the density dependence of the symmetry energy will be attained by
%including $\rho\sigma$ and $\rho\omega$ non-linear terms in the
%Lagrangian density. This will allow us the span the $L$ range  50 to 110 MeV.

% We also calculated
%what happens when we have only baryonic star as a benchmark.
%In all we generated $2\times[9\times(4\times25+50)+25]=2750$ different EOS
%and their star families. %See Table \ref{table:this_work} for reference.
%The effect of $L$ and the meson nonlinear terms are
%thus studied and compared in great detail.
%In all these cases we used only the TM1-2\cite{providencia13} parametrization
%of relativistic mean field models.
%These models are fitted to the ground state properties of a number
%of nuclei and to high density Dirac-Brueckner-Hartree-Fock computations,
%in particular the TM1-2 is a variation of TM1 of \cite{tm1_sumioshi}
%with the difference that the quartic $\omega$ term is smaller than
%the TM1 case thus obtaining harder EOS at large densities.

In section \ref{section:formalism} we present the formalism,
in section \ref{section:results_and_discussion} we show the results,
in section \ref{section:conclusions} we draw some conclusions, and in
appendix  tables with 
several neutron star properties are presented.

\section{FORMALISM}
\label{section:formalism}

In the present section we present a brief review of the model and
discuss the choice of the parameters.
\subsection{EQUATION OF STATE}
%\label{subsection:star_matter} 

For the description of the EOS of neutron star matter, we adopt an RMF
approach in which the nuclear interaction is described by the exchange
of mesons. The baryons considered in this work are nucleons (n and p)
and hyperons ($\Lambda$, $\Sigma$, and $\Xi$). The exchanged mesons
include scalar and vector mesons ($\sigma$ and $\omega$), isovector
meson ($\rho$), and two additional strangeness mesons, the scalar
$\sigma^*$ and the vector $\phi$. The Lagrangian density includes
several nonlinear terms in order to describe adequately the saturation
and high density 
properties of nuclear matter. In the present work, we consider a Lagrangian 
with  nonlinear $\omega$-$\rho$ and $\sigma$-$\rho$ couplings,
characterized by the two coupling constants $\Lambda_\omega$ and $\Lambda_\sigma$
respectively, as introduced in \cite{horowitz2001a,horowitz2001b,carriere2003}.
These couplings allow us to study the effect of the density dependence of the symmetry energy.

Heavy ion collision experiments \cite{heavy_ion_danielewicz} suggest that the high density EOS
should not be too stiff. In the RMF approach these constraints  on the high density EOS
may be implemented including a fourth order term on the vector isoscalar
 meson $\omega$ in the Lagrangian density \cite{tm1,muller1996}, 
or including density dependent couplings \cite{tw1999}. In the present
study we include a fourth order $\omega$ term and  will impose that at
high baryonic density  the EOS satisfies the
constraints proposed in \cite{heavy_ion_danielewicz}, even though theses
constraints should be taken with care since they carry some
uncertainties due to the 
modelling of  heavy ion flow.

For neutron star matter consisting of neutral mixture of baryons and leptons in $\beta$ equilibrium, we start from the effective Lagrangian density of the nonlinear Walecka
model (NLWM) 
\begin{widetext}
\begin{eqnarray}
\label{formula:main_model_hyperon_Lagrangian}
{\cal L}&=&\sum_{B}\bar{\Psi}_{B} \left[\gamma_{\mu}D^{\mu}_{B}- m^{*}_{B}\right]\Psi_{B}+\sum_l\bar{\Psi}_l\left[i\gamma_\mu\partial^\mu-m_l\right]\Psi_l+\frac{1}{2}\left(\partial_{\mu}\sigma \partial^{\mu}\sigma-m^{2}_{\sigma}\sigma^{2}\right)-\frac{\kappa}{3!}\sigma^3-\frac{\lambda}{4!}\sigma^4+\frac{1}{2} m^{2}_{\omega}\omega_{\mu}\omega^{\mu}\cr
&-&\frac{1}{4} \Omega_{\mu \nu} \Omega^{\mu \nu}+\frac{\xi}{4!} g_{\omega}^4 \left(\omega_{\mu}\omega^{\mu}\right)^2+\frac{1}{2}m^{2}_{\rho}\vec{\rho}_{\mu}\cdot\vec{\rho}\,^{\mu}-\frac{1}{4} \vec{R}_{\mu \nu}\cdot \vec{R}^{\mu \nu}+g^{2}_{\rho} \vec{\rho}_{\mu}\cdot\vec{\rho}\,^{\mu}\left[\Lambda_{\omega}g^{2}_{\omega} \omega_{\mu}\omega^{\mu}+\Lambda_{\sigma}g^{2}_{\sigma} \sigma^2\right]\cr
&+&\frac{1}{2}\left(\partial_\mu\sigma^*\partial^\mu\sigma^*-m_{\sigma^*}^2{\sigma^*}^2\right)+\frac{1}{2}m_\phi^2\phi_\mu\phi^\mu-\frac{1}{4}\Phi_{\mu\nu}\Phi^{\mu\nu}
\end{eqnarray}
\end{widetext}
where $D^{\mu}_{B}=i\partial^\mu-g_{\omega_B}\omega^\mu - g_{\phi_B}\phi^\mu-\frac{1}{2}g_{\rho_B}\vec{\tau}_{B}\cdot\vec{\rho}\,^\mu$, and $m_{B}^*=m_{B}-g_{\sigma_B}\sigma-g_{\sigma^*_B}\sigma^*$ is the baryon effective mass, $g_{i_B}$ are the $i=\sigma,\omega,\rho$ meson coupling constants, $\Psi_B$ and $\Psi_l$ are the Dirac fields for the baryons and the leptons, respectively. The baryon mass and the lepton mass are denoted by $m_B$ and $m_l$, respectively. The sum in $B$ is over the eight lightest baryons $n, p, \Lambda, \Sigma^+, \Sigma^0, \Sigma^-, \Xi^0, \Xi^-$. The sum in $l$ is over the two leptons ($e^-$ and $\mu^-$). 
The constants $\kappa$ and $\lambda$ are the couplings of the scalar self-interaction terms and $\tau_B$ is the isospin operator.
The operators {$\Omega_{\mu\nu}=\partial_\mu\omega_\nu-\partial_\nu\omega_\mu$}, {$\Phi_{\mu\nu}=\partial_\mu\phi_\nu-\partial_\nu\phi_\mu$}
and {$\vec{R}_{\mu\nu}=\partial_\mu\vec{\rho}_\nu-\partial_\nu\vec{\rho}_\mu-g_\rho\left(\vec{\rho}_\mu\times\vec{\rho}_\nu\right)$}
are the mesonic field tensors. The hidden-strangeness mesons are represented by the $\sigma^*$ and the $\phi^\mu$ fields. In the RMF approximation, the meson fields are replaced by their expectation
values in the groundstate of the system under consideration. The Euler-Lagrange equations applied to Eq. \eqref{formula:main_model_hyperon_Lagrangian} and using the mean-field approximation, yield the meson field equations  written as
\begin{subequations}
\label{formula:MFA_equations_of_motion_with_hyperons}
\begin{eqnarray}
\label{formula:MFA_equations_of_motion_with_hyperons_sigma}
\sigma_0&=&\frac{1}{m_\sigma^{*2}}\sum_B \frac{g_{\sigma_B}}{\pi^2}\int_0^{k^B_{F}}\frac{m_B^*k^2}{\sqrt{k^2+m_B^{*2}}}\,dk\\
\cr
\label{formula:MFA_equations_of_motion_with_hyperons_omega}
\omega_0&=&\frac{1}{m_\omega^{*2}}
\sum_B \frac{g_{\omega_B}}{3\pi^2}\left(k^B_{F}\right)^3\\
\cr
\label{formula:MFA_equations_of_motion_with_hyperons_rho}
\rho_{03}&=&\frac{1}{m_\rho^{*2}}\sum_B \frac{g_{\rho_B}}{3\pi^2}\tau_{3B}\left(k^B_{F}\right)^3\\
\cr
\label{formula:MFA_equations_of_motion_with_hyperons_sigma*}
\sigma^*_0&=&\frac{1}{m_{\sigma^*}^{2}}\sum_B\frac{g_{\sigma^*_B}}{\pi^2}\int_0^{k^B_{F}}\frac{m_B^*k^2}{\sqrt{k^2+m_B^{*2}}}\,dk\\
\cr
\label{formula:MFA_equations_of_motion_with_hyperons_phi}
\phi_0&=&\frac{1}{m_\phi^{2}}\sum_B \frac{g_{\phi_B}}{3\pi^2}\left(k^B_{F}\right)^3
\end{eqnarray}
\end{subequations}
where $k^B_{F}$ is the $B$ baryon Fermi momentum, and the meson "effective" masses ($m_{i}^{*2},\; i=\sigma,\omega,\rho$) are defined as
\begin{subequations}
\label{meson_effective_masses}
\begin{eqnarray}
\label{meson_effective_masses1}
m_\sigma^{*2}&=&m_\sigma^2+\frac{\kappa}{2}\sigma_0+\frac{\lambda}{6}\sigma_0^2-2\Lambda_{\sigma}g^{2}_{\sigma} g^{2}_{\rho} \rho_{03}^2,\\
\cr
\label{meson_effective_masses2}
m_\omega^{*2}&=&m_\omega^2+\frac{\xi}{6}g_\omega^4\omega_0^2+2\Lambda_{\omega}g^{2}_{\omega} g^{2}_{\rho} \rho_{03}^2,\\
\cr
\label{meson_effective_masses3}
m_\rho^{*2}&=&m_\rho^2+2g^{2}_{\rho} \left[\Lambda_{\omega}g^{2}_{\omega} \omega_0^2+\Lambda_{\sigma}g^{2}_{\sigma} \sigma_0^2\right].
\end{eqnarray}
\end{subequations}
The extra nonlinear term $\sigma$-$\rho$, in the Lagrangian density, causes a decrease of the effective mass
of the $\sigma$-meson with density and the $\omega$-$\rho$ term causes an increase of the $\omega$ meson effective mass. However,
both $\sigma$-$\rho$ and $\omega$-$\rho$ nonlinear terms increase the $\rho$ meson "effective mass" with density.
Consequently, the $\sigma$ meson field will harden at large densities whilst the $\omega$ and $\rho$ fields will soften at large densities. 

For neutron star matter composed by a neutral mixture of baryons and leptons, we force the generalized $\beta$ equilibrium
{ with no} neutrino-trapping and the charge neutrality which can be written as following
\begin{subequations}
\begin{eqnarray}
\label{formula:chemical_equilibrium}
\mu_i-b_i\mu_n&=&q_i\mu_l,\ \mu_S=0\\
\cr
\label{formula:charge_neutrality}
\rho_{p^+}+\rho_{\Sigma^+}&=&\rho_{e^-}+\rho_{\mu^-}+\rho_{\Sigma^-}+\rho_{\Xi^-}
\end{eqnarray}
\end{subequations}
where $q_i$ is the electrical charge, $b_i$ is the baryon number, $\mu_i$ and $\rho_i=(k^i_F)^3/3\pi^2$ are the chemical potential and the number density of species $i$, respectively. The chemical potentials for the baryons ($\mu_B$) and for the leptons ($\mu_l$) are defined by
\begin{subequations}
\begin{eqnarray}
\mu_B&=&\sqrt{\left(k^B_{F}\right)^2+m_B^{*2}}+g_{\omega_B}\omega_0+\frac{g_{\rho_B}}{2}\tau_3\rho_{03}\cr
&+&g_{\phi_B}\phi_0,\\
\cr
\mu_l&=&\sqrt{\left(k^{l}_{F}\right)^2+m_l^{2}}.
\end{eqnarray}
\end{subequations}
The coupled \eqref{formula:MFA_equations_of_motion_with_hyperons_sigma},
\eqref{formula:MFA_equations_of_motion_with_hyperons_omega}, \eqref{formula:MFA_equations_of_motion_with_hyperons_rho},
\eqref{formula:MFA_equations_of_motion_with_hyperons_sigma*}, \eqref{formula:MFA_equations_of_motion_with_hyperons_phi},
\eqref{formula:chemical_equilibrium},
\eqref{formula:charge_neutrality} equations are solved self-consistently
at a given baryon density $\rho_B=\rho_{p}+\rho_{n}+\rho_{\Lambda}+\rho_{\Sigma^+}+\rho_{\Sigma^0}+\rho_{\Sigma^-}
+\rho_{\Xi^0}+\rho_{\Xi^-}$.
From the energy-momentum tensor, we calculate the energy density ($\varepsilon$) and the pressure ($P$):
\begin{widetext}
\begin{eqnarray}
\label{formula:MFA_EOS_energy_density_hyperons}
\varepsilon&=&\sum_B\frac{1}{\pi^2}\int_0^{k^B_{F}}k^2\sqrt{k^2+m_{_B}^{\,_*2}}\,dk+
\sum_l\frac{1}{\pi^2}\int_0^{k^{l}_{F}}k^2\sqrt{k^2+m_l^2}\,dk
+\frac{1}{2}m^{2}_{\sigma}\sigma_0^{2}+\frac{\kappa}{3!}\sigma_0^3+\frac{\lambda}{4!}\sigma_0^4\cr
&+&\frac{1}{2} m^{2}_{\omega}\omega_0^2
+\frac{\xi}{8} g_{\omega}^4 \omega_0^4
+\frac{1}{2}m^{2}_{\rho}\rho_{03}^2
+3g^{2}_{\rho} \rho_{03}^2\left[\Lambda_{\omega}g^{2}_{\omega} \omega_0^2+
\frac{\Lambda_{\sigma}}{3}g^{2}_{\sigma} \sigma_0^2\right]
+\frac{1}{2}m_{\sigma^*}^2{\sigma_0^*}^2
+\frac{1}{2}m_\phi^2\phi_0^2
\end{eqnarray}
\begin{eqnarray}
\label{formula:MFA_EOS_pressure_hyperons}
P&=&\sum_B\frac{1}{3\pi^2}\int_0^{k^B_{F}}\frac{k^2}{\sqrt{k^2+{m_{_B}^*}^2}}\,k^2dk
+\sum_l\frac{1}{3\pi^2}\int_0^{k^{l}_{F}}\frac{k^2}{\sqrt{k^2+m_{_l}^2}}\,k^2dk
-\frac{1}{2}m^{2}_{\sigma}\sigma_0^{2}
-\frac{\kappa}{3!}\sigma_0^3-\frac{\lambda}{4!}\sigma_0^4\cr
&+&\frac{1}{2} m^{2}_{\omega}\omega_0^2
+\frac{1}{4!}\xi g_{\omega}^4 \omega_0^4
+\frac{1}{2}m^{2}_{\rho}\rho_{03}^2
+g^{2}_{\rho} \rho_{03}^2\left[\Lambda_{\omega}g^{2}_{\omega} \omega_0^2+
\Lambda_{\sigma}g^{2}_{\sigma} \sigma_0^2\right]
-\frac{1}{2}m_{\sigma^*}^2{\sigma_0^*}^2
+\frac{1}{2}m_\phi^2\phi_0^2
\end{eqnarray}
\end{widetext}
where  $k^l_{F}$ is the $l$ lepton Fermi momentum.
To calculate neutron star structure {we solve} the general relativity Tolmann-Oppenheimer-Volkov (TOV) pair of equations which are written as
\begin{center}
\begin{subequations}
\label{formula:oppenheimer_volkov_equations}
\begin{eqnarray}
\frac{dP}{dr}(r)=-\frac{P(r)+\varepsilon(r)}{r\left[r-2M(r)\right]}\left\{M(r)+4\pi
  r^3P(r)\right\}
\label{formula:oppenheimer_volkov_equations_pressure_mass_density}\\
M(r)=4\pi\int_0^r\varepsilon(r)r^2dr
\label{formula:oppenheimer_volkov_equations_energy_mass_density}
\end{eqnarray}
\end{subequations}
\end{center}
where $\varepsilon$ is the energy density, $P$ is the pressure and
$M(r)$ is the mass inside radius $r$. The total strangeness content of
a given star, $N_{S}$ is given by
\begin{eqnarray}
\label{formula:total_hyperon_content}
N_{S}=4\pi\int_0^R\frac{r^2}{\sqrt{1-\frac{M(r)}{r}}}\rho_{S}\ dr
\end{eqnarray}
where
\begin{eqnarray}
\label{formula:total_strangeness_density}
\rho_{S}=\rho_{\Lambda}+\rho_{\Sigma^+}+\rho_{\Sigma^0}+\rho_{\Sigma^-}+2\rho_{\Xi^0}+2\rho_{\Xi^-}
\end{eqnarray}
is the total strangeness density.

\subsection{MODEL PARAMETERS}
\label{subsection:model_parameters} 
The RMF models are generally  characterized by a set of nuclear matter
properties at saturation density, including
the nuclear saturation density $\rho_0$, the binding energy per baryon number $B/A$,
the nucleon effective mass $m^*$, the nuclear incompressibility $K_0$,
 the symmetry energy $J$ and its slope $L$. 
In this work we use the TM1-2 parameter set \cite{providencia13}, a modified parametrization  of the TM1 parameter set \cite{tm1}, 
and whose saturation properties are listed in table
\ref{table:tm1_2_saturation_properties}.
At suprasaturation densities the TM1-2 model is slightly stiffer than
the TM1 model but still
overlaps heavy ion flow data \cite{heavy_ion_danielewicz}.

\begin{table}[htb]
\caption{Saturation properties of the TM1-2 model. All units are in MeV.
\label{table:tm1_2_saturation_properties}}
\begin{ruledtabular}
\begin{tabular}{cccccc}
$\rho_0$ (fm$^{-3}$) & $B/A$ & $m^*$ & $K_0$ & $Q_0$    \\
\hline
0.145 & -16.38 & 595.63 & 277.03 & -197.48   \\
\hline
\hline
$\epsilon_\textrm{sym}$ & $L$ & $K_\textrm{sym}$ & $Q_\textrm{sym}$ & $K_\tau$   \\
\hline
36.84 &  111.27 & 41.90 & -32.72  & -546.27
\end{tabular}
\end{ruledtabular}
\end{table}

The parameters of the model are the nucleon mass $m_N=938$ MeV,
  the masses of mesons $m_{\sigma}=511.198$MeV, $m_{\omega}=~783$~MeV,
  $m_{\rho}=770$~MeV, and the coupling constants, which are listed in 
Tables
\ref{table:TM1-2_parametrization}  and \ref{table:sample_values_of_lmbi_gri_pairs_and_L}.

\begin{table}[htb]
\caption{Coupling constants and meson masses for the TM1-2 model.
\label{table:TM1-2_parametrization}}
\begin{ruledtabular}
\begin{tabular}{cccccc}
$(\frac{g_\sigma}{m_\sigma})^2$& $(\frac{g_\omega}{m_\omega})^2$&$(\frac{g_\rho}{m_\rho})^2$&$\kappa/M$&$\lambda$&$\xi$\\
fm$^2$                         &    fm$^2$                      &     fm$^2$                &  \phantom{u}     &    \phantom{u}  &    \phantom{u}  \\
\hline
14.8942                        &    9.9285                      &     5.6363                &   3.52353        &  -47.36246      &  0.01167    \\
\end{tabular}
\end{ruledtabular}
\end{table}

\subsubsection{Density dependence of  symmetry energy}
\begin{figure}[htb]
% figura 2
\includegraphics[width=1.0\linewidth,angle=0]{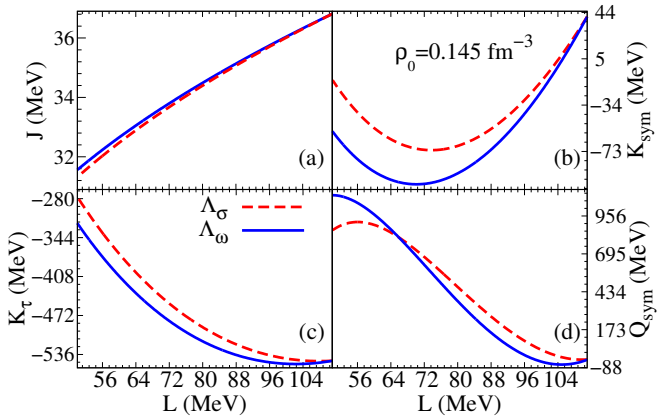}
\caption{
\label{figure:symmetry_tm1-2}
(Colour online) Symmetry energy and derivatives at saturation density
as a function of the slope of the symmetry energy for the TM1-2 model.
}
\end{figure}

%\label{subsection:symmetry_energy_variables} 
The $\omega\rho$ and $\sigma\rho$ nonlinear terms affect the density
dependence of the  symmetry
energy\cite{horowitz2001a,carriere2003,rafael11},
the symmetry energy being given by
\begin{eqnarray}
\label{formula:symmetry_energy_for_symmetric_matter}
\epsilon_\textrm{sym}&=&\frac{k_F^2}{6\sqrt{k_F^2+(m_N-g_\sigma\sigma_0)^2}}\cr
&+&\frac{g_\rho^2}{8}\frac{\rho}{m_\rho^2+2g_\rho^2[\Lambda_\omega(g_\omega\omega_0)^2+\Lambda_\sigma(g_\sigma\sigma_0)^2]}
\end{eqnarray}
$k_F$ is the  symmetric nuclear matter Fermi momentum and $\rho=2k_F^3/3\pi^2$ is the
baryon number density. We have adjusted the symmetry energy to be
$\epsilon_\textrm{sym}=25.52$ MeV at $\rho=0.1\,\textrm{fm}^{-3}$, and taking care that for the minimum possible values of $L$,
the EOS of neutron matter has no binding. We point out that this value
of the symmetry energy at $0.1$ fm$^{-3}$ is inside the range
25.5(10) MeV obtained
in \cite{brown13}  from the properties of doubly magic nuclei. The
range of values covered by $J$ and $L$ are, respectively,
$31.32<J<36.84$ MeV  and $50<L<111$ MeV, see 
Fig. \ref{figure:symmetry_tm1-2} (a). These values cover the values
obtained from isospin diffusion  observables in heavy ion
collisions\cite{tsang09} or the mean N/Z distributions of emitted
fragments with radioactive ion beams \cite{kohley13}, but are not as
low as the ones obtained from chiral effective field theory \cite{kruger13}.

\begin{figure}[htb]
\includegraphics[width=1.0\linewidth,angle=0]{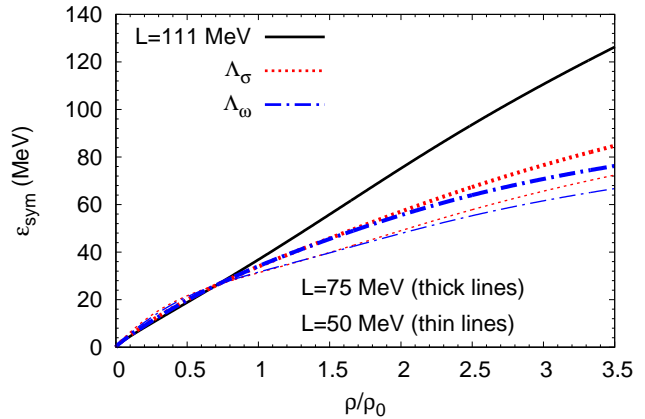}
\caption{
\label{figure:comparing_esym_lmbs_and_lmbw}
(Color online) Symmetry energy versus density for the TM1-2 parameter set and
comparison between the $\omega\rho$ and $\sigma\rho$ behaviors for
$L=75$ MeV (thick colored lines) and $L=50$ MeV (thin colored lines).
}
\end{figure}

\begin{table}[htb]
\caption[Sigma couplings]{
\label{table:sample_values_of_lmbi_gri_pairs_and_L}
Values of some of the $\Lambda_i,g_{\rho_i}$ pairs and their correspondent symmetry energy slopes.
}
\begin{ruledtabular}
\begin{tabular}{cccc}
$L$ (MeV)       &                       &       & $g_\rho$ \\
\hline
111             &  $\Lambda_i$          & 0.00  & 9.26     \\
\hline
75              &  $\Lambda_\sigma$     & 0.01  & 10.42    \\
\phantom{u}     &  $\Lambda_\omega$     & 0.02  & 10.18    \\
\hline
50              &  $\Lambda_\sigma$     & 0.02  & 12.76    \\
\phantom{u}     &  $\Lambda_\omega$     & 0.03  & 11.78    \\
\end{tabular}
\end{ruledtabular}
\end{table}
In Fig. \ref{figure:comparing_esym_lmbs_and_lmbw}, see also  Table \ref{table:sample_values_of_lmbi_gri_pairs_and_L}
 for the parameters $\Lambda_i,g_{\rho_i}$ used,  
we plot the symmetry
energy as a function of the density for three values of $L$, 111 MeV,
corresponding to the model with no $\omega\rho$ nor $\omega\sigma$
terms, 75 and 50 MeV. 
Below $\rho=0.1$ fm$^{-3}$, where all curves cross, the $\rho\sigma$ term has a stronger
effect, while above due to the saturation of the $\sigma$-field the
$\rho\omega$ term gives rise to a stronger softening. We expect, therefore,
 a stronger effect of the non-linear term $\rho\sigma$ on the
properties of the crust while the $\rho\omega$ term will have a larger
influence on properties determined by the suprasaturation EOS.

As reference, we plot  in Fig. \ref{figure:symmetry_tm1-2}  the 
symmetry energy, the incompressibility $K_{sym}$,  the incompressibility
coefficient $K_\tau$ and the third derivative of the
symmetry energy $Q_{sym}$,  all defined at saturation, as a
function of $L$. It is clearly seen  
the existence of correlations between the parameters that define the isovector
channel, and in general these correlations  agree with the correlations
discussed in \cite{vidana09,camille11}. The non linear terms
$\omega\rho$ and $\sigma\rho$ give similar results, although with the
$\sigma\rho$ non-linear term a steeper behavior with $L$ is obtained for both $J$
and $K_\tau$.

\subsubsection{Hyperon couplings}
%\label{subsection:hyperons}

 The NLWM extended lagrangian density includes also the mass of the eight baryons of the
baryonic octet, the mass of $\sigma^*$ and the $\phi$ mesons, the electron and muon
masses and the meson-hyperon couplings that we will discuss next.

The $\omega$ and $\phi$ isoscalar vector mesons coupling constants $g_{\omega_B}$ and $g_{\phi_B}$ are
determined by SU(6) symmetry:
\begin{subequations}
\begin{eqnarray}
\label{formula:hyperon_coupling_constants_omega}
\frac{1}{3}g_{\omega_N}=\frac{1}{2}g_{\omega_\Lambda}=\frac{1}{2}g_{\omega_\Sigma}=g_{\omega_\Xi}\\
\cr
\label{formula:hyperon_coupling_constants_phi}
2g_{\phi_\Lambda}=2g_{\phi_\Sigma}=g_{\phi_\Xi}=-\frac{2\sqrt{2}}{3}g_{\omega_N}.
\end{eqnarray}
\end{subequations}
For the $\rho$-meson we take
\begin{equation}
\label{formula:hyperon_coupling_constants_rho}
g_{\rho_N}=g_{\rho_\Sigma}=g_{\rho_\Xi}, \, g_{\rho_\Lambda}=0\\
\end{equation}
and the isospin operator in the baryon-meson coupling term in the
Lagrangian takes into account the isospin symmetry.
To define the hyperon-$\sigma$ meson couplings $g_{\sigma_B}$, 
we use the hyperonic potential in symmetric  nuclear matter at saturation defined by the relation
\begin{eqnarray}
\label{formula:optical_potentials}
U^N_H=x_{\omega H}V_\omega-x_{\sigma H} V_\sigma
\end{eqnarray}
where $V_\omega=g_\omega\omega_0=273.88$~MeV,
$V_\sigma=g_\sigma\sigma_0=342.71$~MeV are defined at saturation density,  $g_{\sigma_H}=x_{\sigma H}g_{\sigma_N}$ and
$g_{\omega_H}=x_{\omega H}g_{\omega_N}$.
While $U^N_\Lambda=-28$ MeV is quite well constrained,  $U^N_\Sigma$
and  $U^N_\Xi$ are not so well constrained:
the experimental data \cite{avraham_gal_hyperon_optical_potentials_2010} suggest that 
$U^N_\Xi$ is attractive, while  $U^N_\Sigma$ is repulsive.
In this work we established $U^N_\Lambda=-28$ MeV for all calculations
and we use $U^N_\Sigma=+30,0,-30$ MeV, and  $U^N_\Xi=+18,0,-18$ MeV
to take into account uncertainties.
In  Table \ref{table:optical_potentials} the optical
potential sets (OPS) used in the present work are identified by numbers.

\begin{table}[htb]
\caption
{Reference table for the optical potentials sets used in the study. All sets satisfy
{$U^N_\Lambda=-28$ MeV}. The optical potentials $U_Y^N$ are defined in
symmetric nuclear matter at saturation $\rho_0$.  In the calculations which
include the $\sigma^*$ meson the values listed below for the strange
$\sigma^*$ meson couplings have been used. In all cases we have
$g_{\sigma^*_\Lambda}=8.524$. In the last two lines the values of
$YY$ optical potentials $U_Y^Y$ at $\rho_0/5$ are given, where
in all cases we have $U_\Lambda^\Lambda(\rho_0/5)=-9.83$ MeV. The potentials units are MeV.}
\label{table:optical_potentials}
\begin{ruledtabular}
\begin{tabular}{ccccccccccc}
Set           & 1 & 2 & 3 & 4 & 5 & 6 & 7 & 8 & 9\\
\hline
$U^N_\Sigma$  & +30 & +30 & +30 & 0 & 0 & 0 & -30 & -30 & -30\\
$U^N_\Xi$     & +18 & 0 & -18 & +18 & 0 & -18 & +18 & 0 & -18\\
$g_{\sigma^*_\Sigma}$   & 9.87  & 9.87   & 9.87 & 8.38  & 8.38 & 8.38 & 6.10  & 6.10 & 6.10\\
$g_{\sigma^*_\Xi}$      & 13.01 & 12.68  & 12.27 & 13.01 & 12.68  & 12.27  & 13.01  & 12.68 & 12.27\\
$U_\Sigma^\Sigma$   & -15.75  & -15.74   & -15.74  & -8.80  & -8.80 & -8.80 & -1.81  & -1.81 & -1.81\\
$U_\Xi^\Xi$         & -9.57   & -7.37    & -5.24   & -9.57  & -7.37 & -5.24 & -9.57  & -7.37 & -5.24 \\
\end{tabular}
\end{ruledtabular}
\end{table}

\section{RESULTS AND DISCUSSION}
\label{section:results_and_discussion}

In the following we discuss the properties of the equations of state
and neutron stars obtained spanning the parameters presented in the
last section.

\subsection{Equation of state of hyperonic stars}
%\label{subsection:equation_of_state}

In the present section we  generalize the studies developed in
\cite{rafael11,providencia13,providencia14} and  discuss the influence of the density
dependence of the  symmetry energy on the hyperonic content of a
neutron star. It {has} been shown in \cite{rafael11,providencia13}
that star properties such as the radius, mass, strangeness content or
the central baryonic density depend non-linearly on the slope $L$.

\begin{figure}[htb]
% figura 3
\includegraphics[width=1.0\linewidth,angle=0]{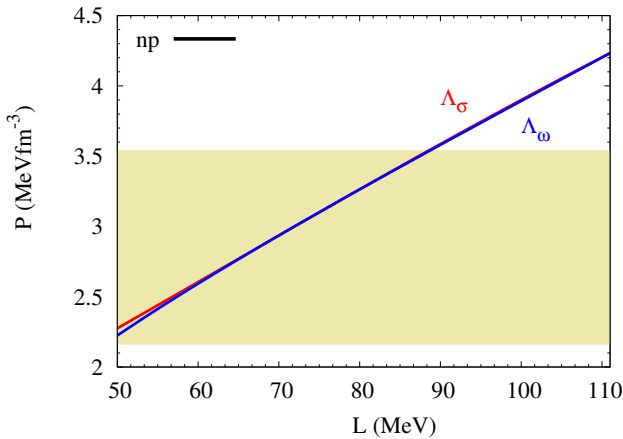}
\caption{
\label{figure:pn0}
(Colour online) Stellar matter pressure at the saturation nuclear
density $\rho_0$. The band defines the range of allowed values calculated in
\cite{hebeler13} from a microscopic calculation.
}
\end{figure}

To make a systematic study
we cover the range $50<L<111$ MeV, for both non-linear terms
$\rho\omega$ and $\rho\sigma$,
generating
an EOS and the corresponding family of neutron stars for each value of
$L$. In Fig. \ref{figure:pn0} we show the stellar matter pressure at the saturation nuclear
density, $P(\rho_0)$, as a function of the slope $L$. The band, that restricts $L$ to
$L<88$ MeV, identifies the  allowed values of $P(\rho_0)$ obtained in
\cite{hebeler13} from a microscopic calculation. In the following we will discuss the strangeness content in the star covering the whole range $50<L<111$ MeV.

\begin{center}
\begin{figure}
\begin{tabular}{c}
% figura 4
\includegraphics[width=0.9\linewidth,angle=0]{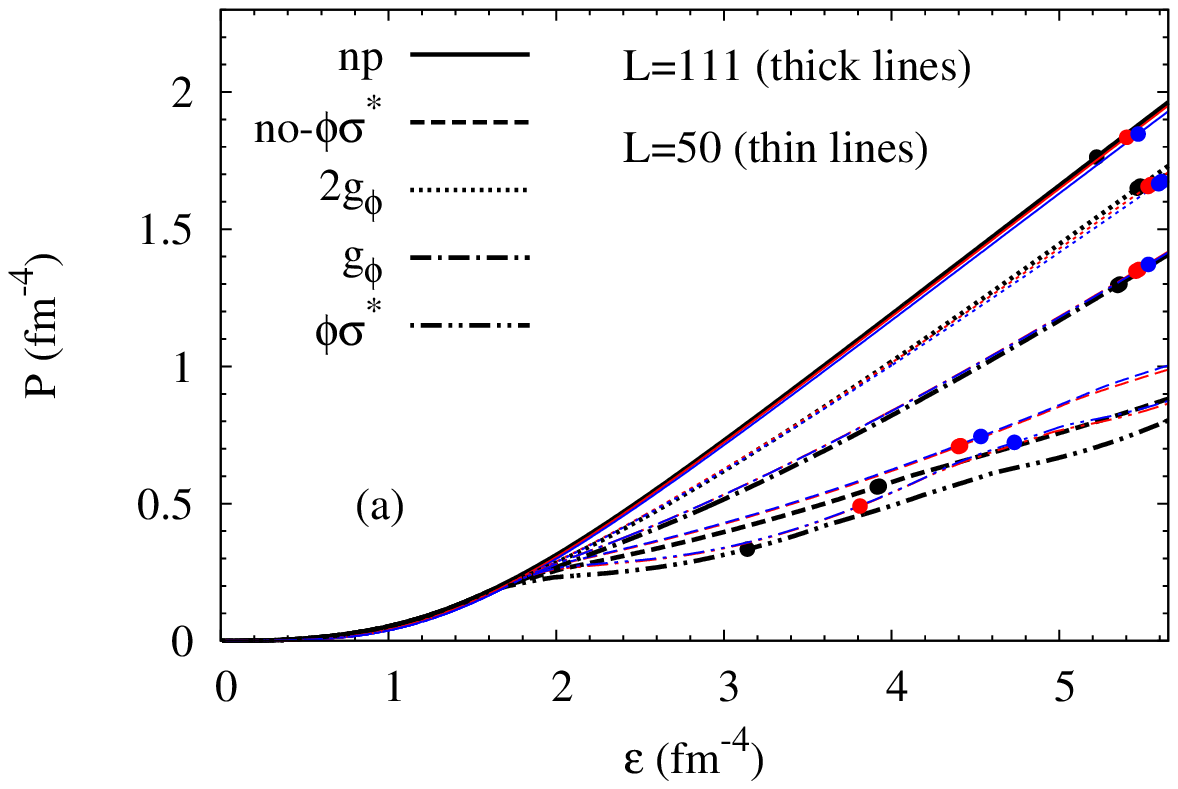}\\
\includegraphics[width=0.9\linewidth,angle=0]{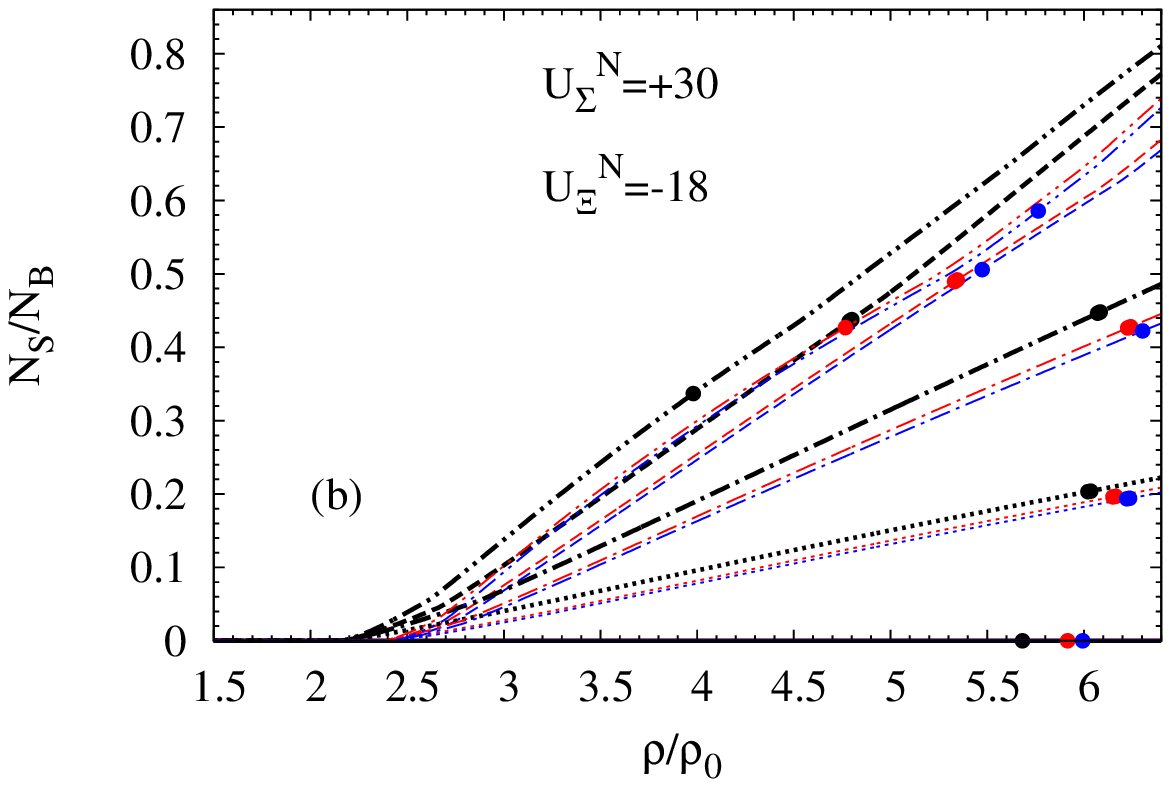}\\
\includegraphics[width=0.9\linewidth,angle=0]{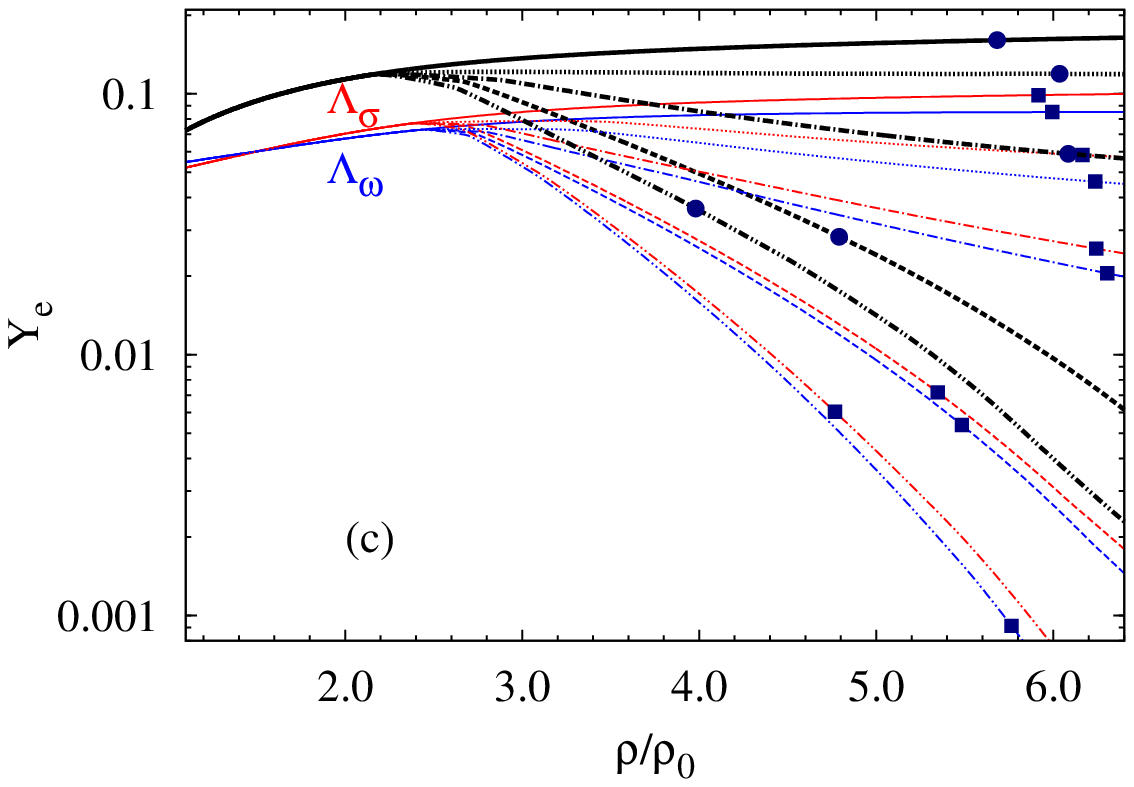}
\end{tabular}
\caption{
(Colour online) Equation of state (top),  strangeness
fraction (middle) and electron fraction (bottom)  for $pn$ matter and different choices of the 
hyperon potentials for the TM1-2 model. 
All figures have been done for  $U^N_\Sigma=+30$ MeV and $U^N_\Xi=-18$ MeV.
The thick black lines correspond to $L=111$ MeV and the thin colored lines correspond to $L=50$ MeV.
Dots indicate the central energy density or baryonic density of the
maximum mass star.}
\label{figure:eos_panel}
\end{figure}
\end{center}

We consider  hyperon-hyperon interactions described by
the exchange of the vector meson $\phi$ and the scalar meson
$\sigma^*$, both with hidden strangeness.
The $\Lambda-\Lambda$ interaction was shown in \cite{gal2011} to be
weakly attractive. This is  implemented in RMF models by
considering a small $g_{\sigma^*_B}$ or even  $g_{\sigma^*_B}=0$
as described in Section \ref{subsection:model_parameters}.

We have considered  five different sets
of coupling constants for the baryonic degrees of freedom in our nuclear matter EOS:
a) one containing only neutrons, protons, electrons and muons in chemical equilibrium which is designated
by $np$; 
b) in a second scenario, designated by
$no$-$\phi\sigma^*$, hyperons are also included.
The hyperon coupling constants to
 the vector mesons $\omega$ and $\rho$ are defined by
Eqs. \eqref{formula:hyperon_coupling_constants_omega} and
\eqref{formula:hyperon_coupling_constants_rho}, and the scalar meson $\sigma$ couplings are defined
by Eq. \eqref{formula:optical_potentials};
 c) a third scenario
excludes the $\sigma^*$ but includes the exchange of the $\phi$ meson  whose coupling constants are defined by
the Eq. \eqref{formula:hyperon_coupling_constants_phi}
with the SU(6) prescription. This case is designated by $g_\phi$ and
takes into account that the recent results seem to indicate that the
binding $\Lambda-\Lambda$ is very weak; 
d)
still keeping only the vector meson $\phi$ but lifting the SU(6)
symmetry,  the $\phi$-hyperon couplings are defined by
\begin{eqnarray}
\label{formula:hyperon_coupling_constants_2phi}
2g_{\phi_\Lambda}=2g_{\phi_\Sigma}=g_{\phi_\Xi}=-\frac{4\sqrt{2}}{3}g_{\omega_N}
\end{eqnarray}
This choice of couplings, designated by $2g_\phi$, brings extra
repulsion between hyperons, affecting mostly the $\Xi$-hyperon;
 e) a last
possibility includes both the $\sigma^*$ and $\phi$ mesons, keeping
SU(6) symmetry to define the hyperon couplings to $\phi$. The
$g_{\sigma^*_B}$ couplings that have been used are listed in
Table \ref{table:optical_potentials}. This case is designated as $\phi\sigma^*$.
The values chosen for
$g_{\sigma^*}^Y$ and listed in Table \ref{table:optical_potentials}
are such that taking for the $\phi$ the SU(6) values the respective
symmetric hyperon matter attains saturation for a binding energy of $1$ MeV.
 The optical
hyperon potentials $U^Y_Y$  of the respective hyperon multiplet
symmetric matter,
\begin{eqnarray}
\label{formula:hyperon_potential_depth}
U^H_H&=&-g_{\sigma_H}\sigma_0-g_{\sigma^*_H}\sigma_0^*+g_{\omega_H}\omega_0+g_{\phi_H}\phi_0,
\end{eqnarray} 
 calculated at $\rho=\rho_0/5$ are also listed in \ref{table:optical_potentials}.

The stellar matter EOS is obtained taking at low densities, below
neutron drip,  the
Baym-Pethick-Sutherland \cite{bps} 
EOS, which is adequate
to describe the neutron star outer crust, 
between neutron drip and 0.01 fm$^{-3}$ Bethe-Baym-Pethick EOS \cite{bps}  is
considered, and above this density the EOS is
smoothly interpolate to the
homogeneous equation of state which is taken above $0.5\rho_0$.

\begin{table}[htb]
\caption{
\label{table:max_mass_no}
Nucleonic maximum mass star properties. 
}
\begin{ruledtabular}
\begin{tabular}{c c c ccccccc}
            &  $L$      & $M_g$  &   $M_b$    &   $R$      &$\varepsilon_0$ &$\rho_c$\\
\hline
                &  111      & 2.27   &    2.67    &   12.50    &   5.216    & 0.82\\

$\Lambda_\omega$ &  50       & 2.22   &    2.63    &   12.02    &   5.474        & 0.87\\
$\Lambda_\sigma$ &  50       & 2.24   &    2.65    &   12.09    &   5.399       & 0.86\\
\end{tabular}
\end{ruledtabular}
\end{table}

In Fig. \ref{figure:eos_panel} we show the EOS, the strangeness fraction and  the electron fraction
for  $L=50$ and 111 MeV and the different choices of the couplings for the baryon degrees of
freedom described above
$np$, $no$-$\phi\sigma^*$, $g_\phi$, $2g_\phi$ and
$\phi\sigma^*$.
The dots indicate the central energy or baryonic  densities corresponding to the maximum mass neutron star.

Some general conclusions, see also \cite{rafael11,schaffner12b,providencia13}, can be drawn:
a) both the strangeness  and a small symmetry energy
slope $L$ give rise to a softer EOS. The  inclusion of
$\phi$ excluding $\sigma^*$ will produce a softer EOS than the $pn$-EOS
but may still be quite
hard if we do not restrict the choice of the couplings to the
SU(6) symmetry; b) the inclusion of strangeness softens more
an EOS with a large than with a small $L$ because the fraction of
hyperons grows faster with density.  For the same value of $L$, the  $\sigma\rho$ non-linear term
favors more the appearance of hyperons than the $\omega\rho$ non-linear
term; c)  including only the  $\phi$-meson and
excluding the $\sigma^*$ reduces the fraction of hyperons due to the extra
repulsion between hyperons introduced. If the
 $\sigma^*$  is also included, the choice of the couplings will determine the behavior
of the EOS and may soften the EOS more than when no strange meson is
included; d) a larger amount of hyperons and a smaller $L$ induce a smaller
amount of electrons, and this effect is stronger for the EOS with the
$\omega\rho$ term.
These main features will justify the results we discuss next for the
dependence of the neutron star properties on the slope $L$ and
strangeness content.

\subsection{Dependence of  hyperonic  star properties on $L$}
%\label{subsection:equation_of_state} 

\begin{figure}[htb]
% figura 5
\includegraphics[width=1.0\linewidth,angle=0]{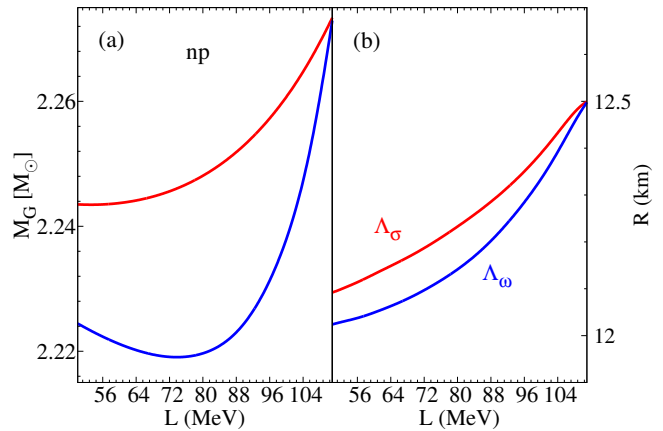}\\
\caption{
(Colour online) Maximum gravitational mass $(a)$,  maximum mass star radius $(b)$
versus $L$ for $np$ matter in $\beta$ equilibrium.
\label{figure:max_mass_PN}}
\end{figure}

\begin{figure}[htb]
% figura 6
\includegraphics[width=1.0\linewidth,angle=0]{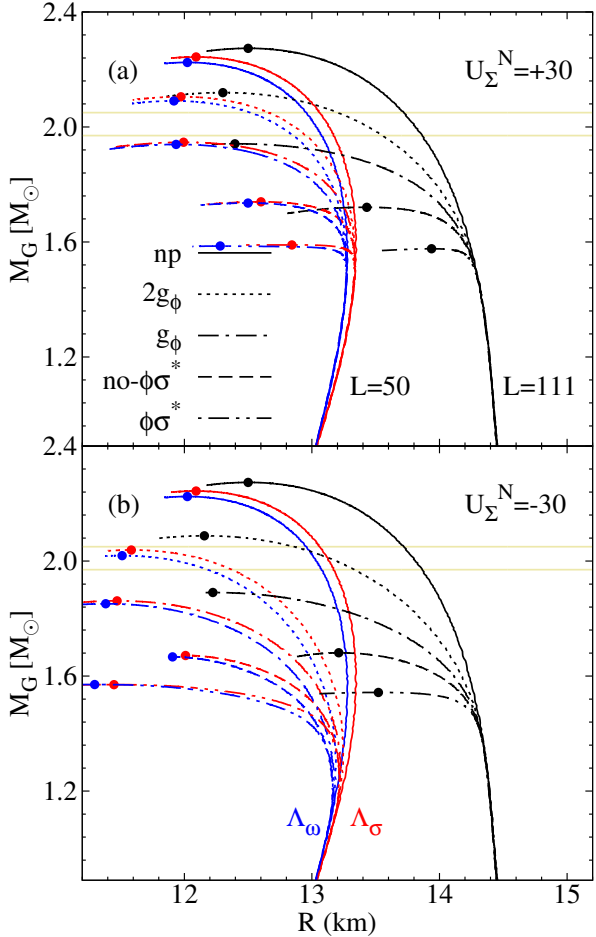}
\caption{
(Colour online) Gravitational mass vs the radius for  two different values of
$L$ with the parameters corresponding to set 3 ($U_\Sigma^N=+30$ MeV,
top)  and set  7  ($U_\Sigma^N=-30$ MeV,
bottom).
The dots indicate the maximum gravitational mass stars.
The faded lines indicate the lower and upper
limits of the pulsar PSR J0348+0432 mass.
}
\label{figure:MR}
\end{figure}

The results obtained
after integrating the TOV equations, Eqs. \eqref{formula:oppenheimer_volkov_equations}, for
all the EOS defined above are presented and discussed in the present subsection.  In Table \ref{table:max_mass_no}  and
Fig. \ref{figure:max_mass_PN}, the
maximum mass star properties obtained with an EOS excluding
strangeness are given. The RMF model used allows maximum masses well above the
limit imposed by the pulsar PSR J0348+0432 mass with $M=2.01\pm
{0.04}\, M_\odot$. Changing the symmetry energy slope, softens the EOS and as a result
the central density is larger and the radius smaller. For the same
value of $L$ this effect is larger when the non-linear term
$\omega\rho$ is used. Similar effect was discussed in
\cite{carriere2003}. The density dependence of the symmetry energy 
affects the maximum star mass only slightly: not more than 1\% with
the $\sigma\rho$ term, and $\lesssim 3\%$ with the $\omega\rho$
 term. In the last case the minimum mass is obtained for $L\sim 80$
 MeV, while in the first the lowest value of $L$ gives the smallest
 mass. The behavior obtained with the $\omega\rho$ term is due to the
 interplay between a softening created for large $L$ due to the larger
 amount of hyperons, and for small values of $L$ due the smaller
 symmetry energy at large densities.

In Fig. \ref{figure:MR} we plot the families of stars obtained
with the parametrizations 3 and 7 of Table \ref{table:optical_potentials} and
consider the different possibilities of including or not including $\phi$ and $\sigma^*$ mesons. For
reference we also include the nucleonic stars.
The dots mark the maximum mass configurations. The main conclusions
drawn from the figure are effects already known: the presence of
strangeness may have a strong effect on the maximum mass
configuration, in particular,  a strong reduction occurs if the $\phi$ meson is not
included, or if it is included together with $\sigma^*$ corresponding
to a quite attractive YY
potential, see, however, \cite{oertel2014} where it  was shown that if
a strong
enough $g_\phi$ coupling is chosen, it was possible to describe a 2$M_\odot$ star
including $\sigma^*$ and $\phi$ mesons;
 a small value of $L$ will give rise to stars with a
smaller radius, the effect being larger if the $\omega\rho$ non-linear
term is used. These general features are, however, sensitive to the
choice of the hyperon optical potentials in symmetric nuclear matter
as we will discuss next. Finally, strangeness is present in less
massive stars if we take an attractive $U_\Sigma^N$ potential.

\begin{figure}[htb]
% figura 7
\includegraphics[width=1.0\linewidth,angle=0]{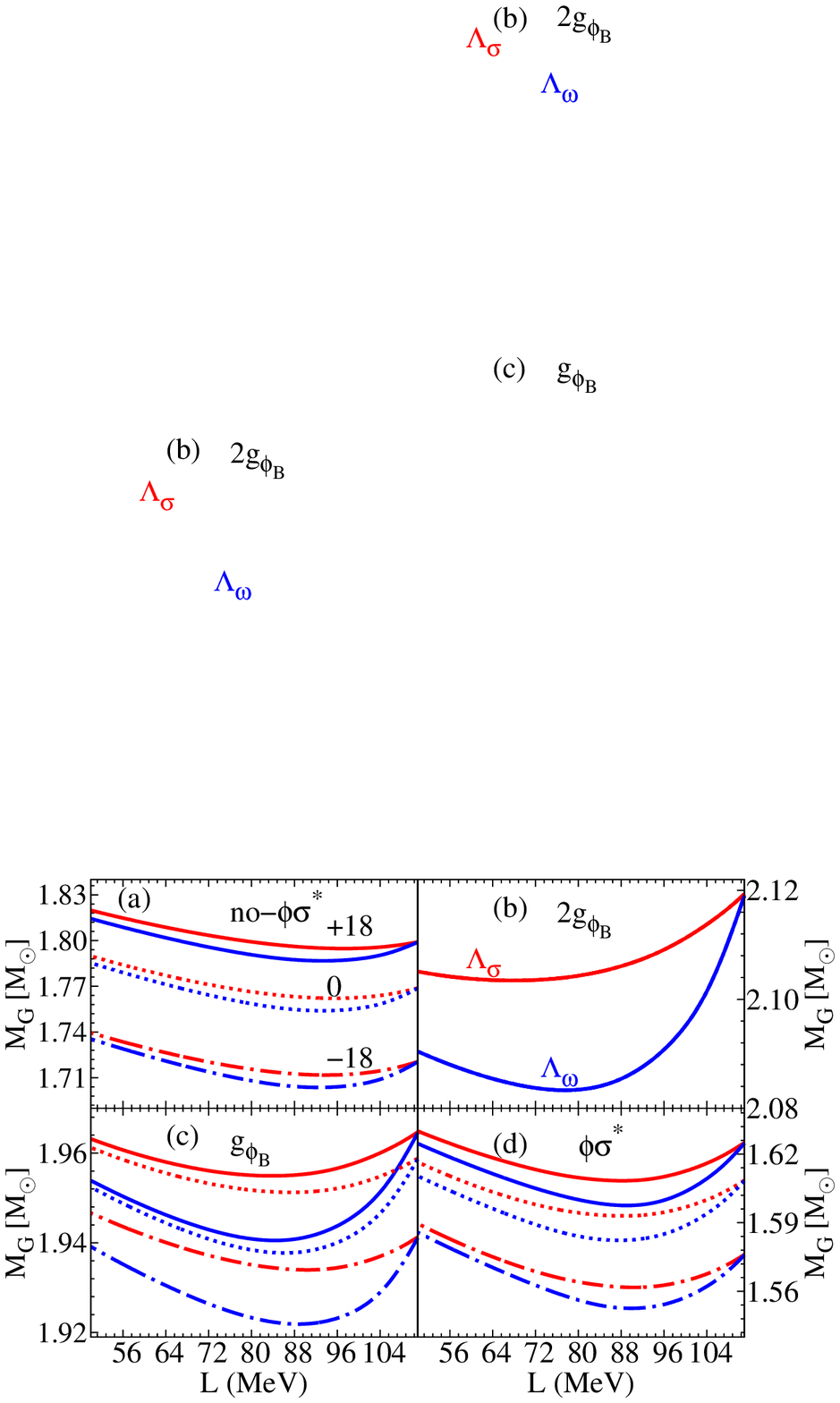}\\
\includegraphics[width=1.0\linewidth,angle=0]{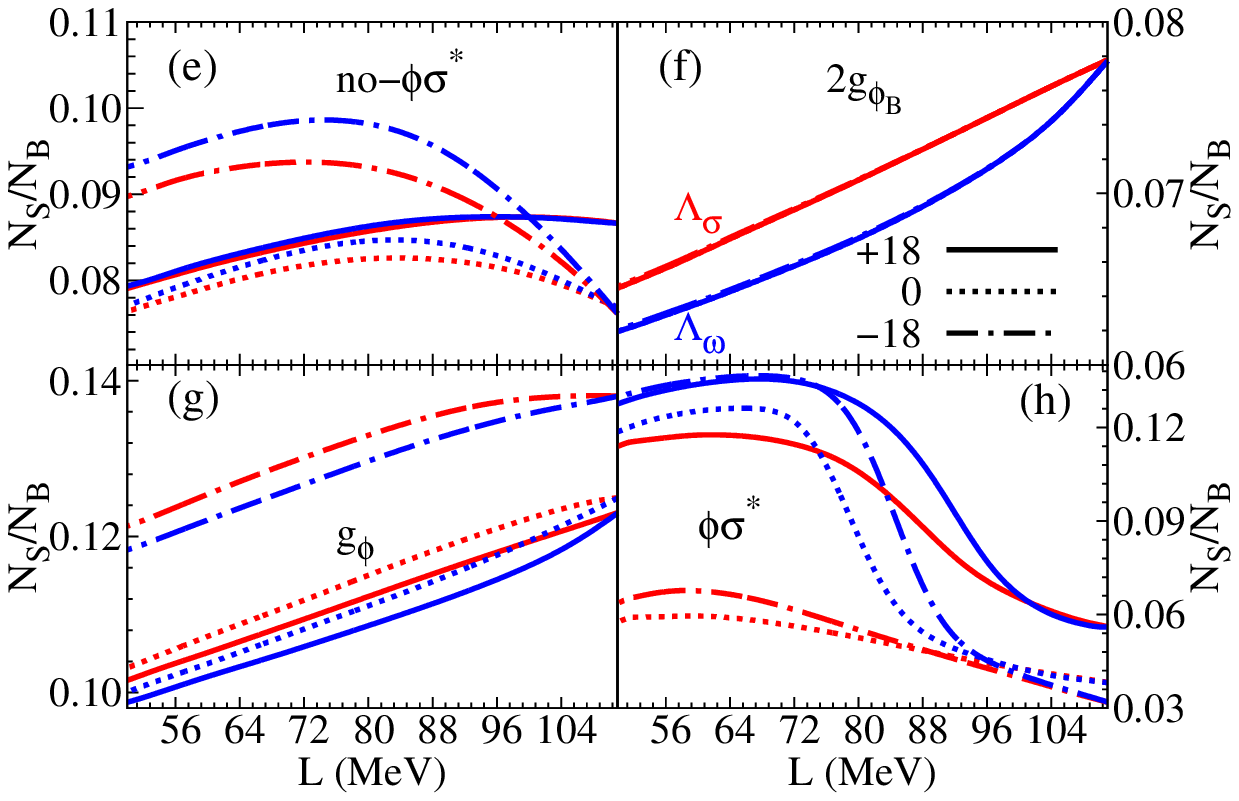}\\
\caption{
(Colour online) Maximum gravitational mass %
and  maximum mass star strangeness fraction versus $L$ for $U_\Sigma^N=+30$ MeV and
all the other optical potentials (+18,0,-18) for $U_\Xi^N$ in MeV.}
\label{figure:max_mass}
\end{figure}

Properties of maximum mass hyperonic stars are given, in appendix,  in Tables
\ref{table:hyp1}, \ref{table:hyp2} and \ref{table:hyp3} corresponding
respectively to stars obtained with $U_\Sigma=+30,\, 0, \, -30$
MeV. Since there is a non-monotonic behavior of the maximum mass with
$L$,  in all cases we give information with respect to stars obtained
with an EOS with $L$=110 MeV, 50 MeV and the value of $L$ that gives
the star with the minimum maximum mass when this does not occur for
$L=50$ MeV. The mass $M_S$ is the mass of
a star with a droplet of  strangeness in its center, \textit{i.e.} it corresponds
to the mass of the star that defines the onset of strangeness in the
star core.

\begin{figure}[htb]
% figura 8
\includegraphics[width=1.0\linewidth,angle=0]{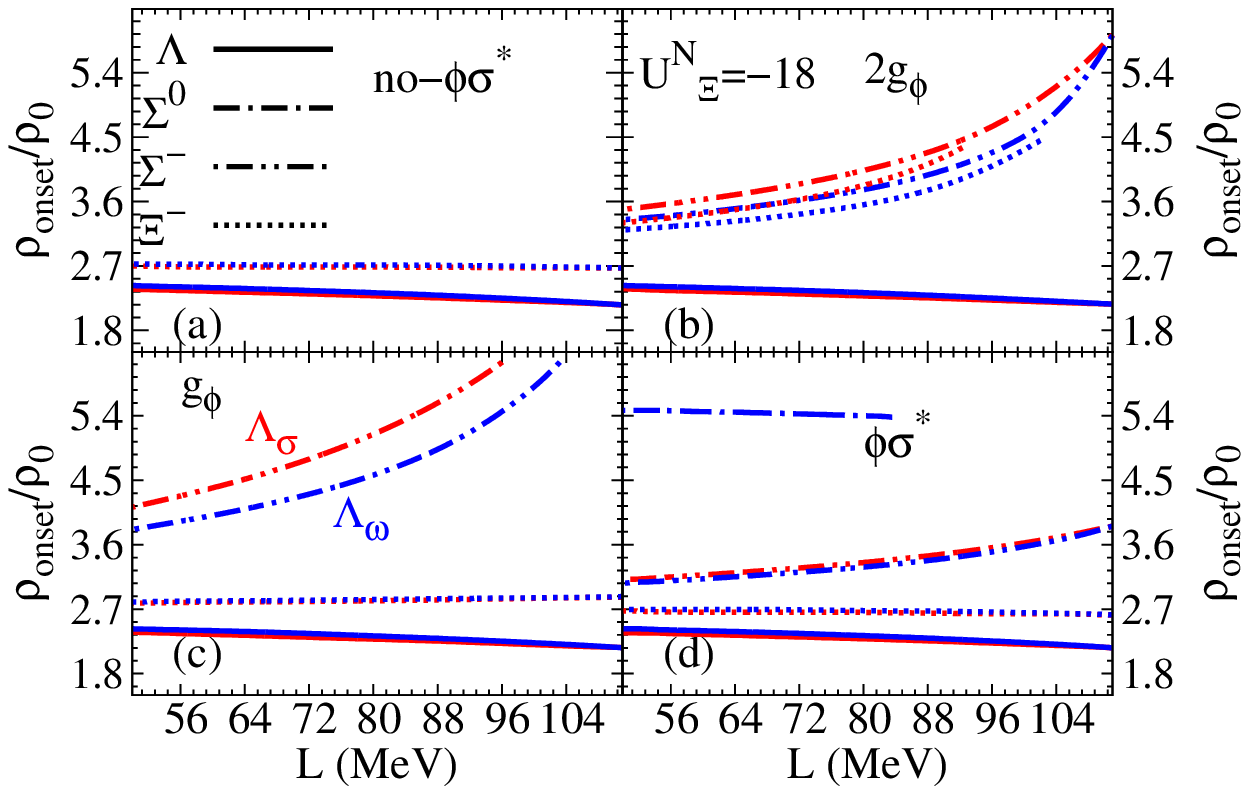}\\
\includegraphics[width=1.0\linewidth,angle=0]{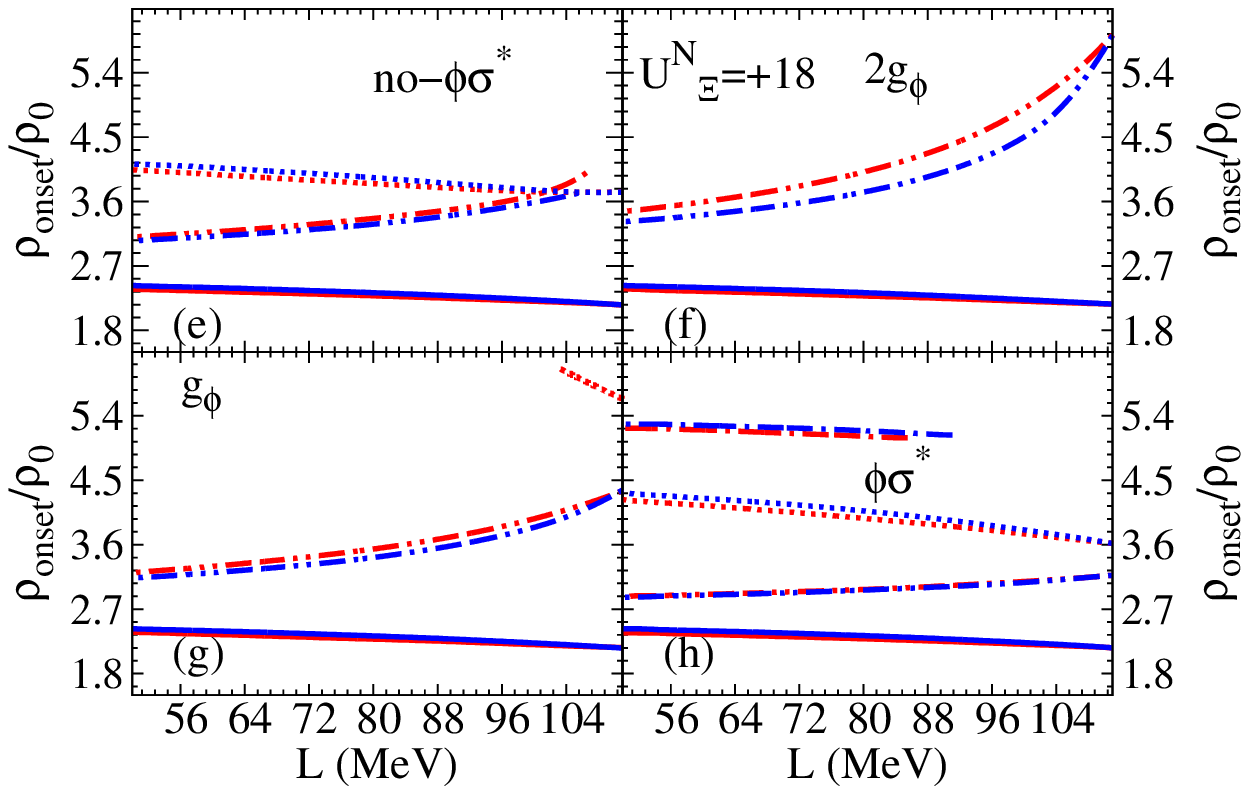}
\caption{
\label{figure:onset}
(Colour online) Onset particle density versus $L$  with $U_\Sigma^N=+30$ MeV (both panels) 
and $U_\Xi^N=-18$ MeV in the top panel and $U_\Xi^N=+18$ MeV in the bottom panel.
}
\end{figure}

In  the two panels of Fig. \ref{figure:max_mass} we plot,
respectively, the gravitational mass (top) and the
strangeness fraction (bottom) of the maximum mass star as a function
of the slope $L$, taking 
$U_\Sigma^N=+30$ MeV.
Results for hyperonic stars are shown for the $U_\Xi^N=+18,0,-18$
MeV, the  nonlinear terms $\sigma\rho$ and $\omega\rho$, and different
choices of the couplings to the $\phi$ and $\sigma^*$ mesons.

All curves  of  plots $(b)$ and  $(c)$  of the top panel  fall within the mass
limits ($M_G=1.97\pm0.04\,M_\odot$) of the observed pulsar PSR
J1614-2230 \cite{197_neutron_star_demorest}
for any $U^N_\Xi$ with the nonlinear $\sigma\rho$ term
and almost all for the nonlinear  $\omega\rho$ term, the exception being
$U_\Xi^N=-18$ MeV  in plot c) in a small interval near the mass minimum.
However, the mass of the pulsar PSR J0348+0432\cite{antoniadis2013},
$M_G=2.01\pm 0.04\,M_\odot$ , is only attained with a stronger
$g_\phi$ than the one obtained imposing SU(6) symmetry. Excluding the
$\phi$ and $\sigma^*$ mesons, or choosing a too attractive YY
potential will not allow hyperonic stars to have a  mass of the order
of 2 $M_\odot$. The maximum
masses obtained with the  attractive YY we have chosen are even
smaller than the ones obtained excluding $\phi$ and $\sigma^*$.
We should point out that maximum mass configurations do not depend on
$U_\Xi^N$ when the $g_\phi$ is very strong because the $\Xi$ hyperon is
disfavored due to its large strangeness charge and the repulsive
effect of the $\phi$-meson. 

Another conclusion that is drawn from the
top panel of Fig. \ref{figure:max_mass} is  that the maximum mass has a non-monotonic
behavior with $L$ and generally it has a  minimum for an intermediate value
of $L$, just as was already seen for nucleonic stars in Fig. \ref{figure:max_mass_PN}.
A small $L$ may give a larger maximum mass due to a smaller
strangeness content. This is the case when $\sigma^*$ is excluded and
$g_\phi$ is defined by the SU(6) symmetry, see panels (c) and (g) in Fig.~\ref{figure:max_mass}.

The total strangeness fraction
evolves nonlinearly with $L$, see 
Fig. \ref{figure:max_mass} bottom panel, and may even show a
non-monotonic behavior with $L$ if no YY or an attractive YY
interaction is chosen, figures a) and d). The large increase of the strangeness fraction
for small values of $L$ in bottom panel d) is due to the onset of the
$\Sigma^0$ hyperon as is clearly seen in panels (d) and (h) of Fig. \ref{figure:onset} obtained respectively for
$U_\Xi^N=-18$ and +18 MeV. 
An attractive $U_\Xi^N$ favors the onset of the $\Xi^-$ hyperon which
is the second hyperon to set in, see top panel. However,  if $g_\phi$ is too
repulsive $\Xi^-$ comes close with the $\Sigma^-$, see (b) top
panel.  A repulsive $U_\Xi^N$ together with a repulsive $YY$
interaction will hinder completely the appearance of the  $\Xi^-$, see
(f) and (g) of bottom panel. 
Fig. \ref{figure:onset} also shows how the
slope $L$ influences the onset of the neutral versus negatively
charged hyperons: the onset density of neutral hyperons decreases with
increasing $L$, while an opposite behavior occurs for the
negatively charges hyperons. A similar conclusion was drawn in \cite{providencia13}.

As a general trend the larger the strangeness fraction the smaller the mass. 
This trend is broken if the $g_\phi$ coupling is very strong and is also influenced 
by the behavior of the symmetry energy with the density. An asy-soft EOS will disfavor
the strangeness onset if $U_\Sigma$ is repulsive: in all plots of both
panels of Fig. \ref{figure:onset}, the density of onset of the
$\Lambda$-hyperon decreases when $L$ increases. However, if the YY interaction is
not too repulsive, intermediate values of $L$  favor the appearance
of negatively charged hyperons that decrease the electron fraction and
soften the EOS, see Fig \ref{figure:eos_panel}. Further decreasing $L$
shifts the hyperon onset to larger
densities, which occurs with the $\Lambda$ onset,  and these effect is
not compensated by the earlier onset of the negatively charged hyperons.

\begin{figure}[htb]
% figura 9
\includegraphics[width=1.0\linewidth,angle=0]{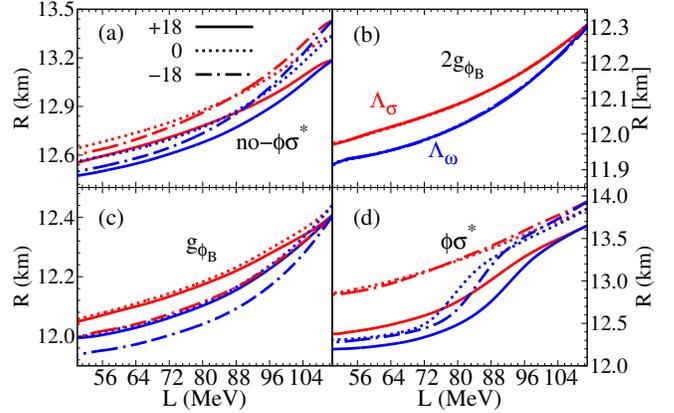}\\
\caption{
(Colour online) Maximum mass star radius
versus $L$ for TM1-2 with $U_\Sigma^N=+30$ MeV and
all the other optical potentials (+18,0,-18) for $U_\Xi^N$ in MeV.}
\label{figure:max_radius}
\end{figure}

In Fig. \ref{figure:max_radius} the behavior of the
maximum mass star radius with $L$ is compared for the different hyperon
parametrizations considered taking $U^N_\Sigma=+30$ MeV. The general trend is a decrease of the
radius if $L$ decreases. This radius reduction may be as large as 1 to 1.5 km if
no YY or a quite attractive YY interaction is considered.

For the parametrization labelled $\phi\sigma^*$ % and all  $L$ values 
the hyperons $\Lambda$, $\Sigma^-$ and
$\Xi^-$ are present in the core independently of $L$. 
Let us now consider  $U^N_\Xi=+18$ MeV, which gives the smallest radii
and has the largest fractions of strangeness with the nonlinear term
$\omega\rho$. For $L\sim 90 - 95$ MeV the curves for the radius and strangeness
fraction suffer a faster change, respectively increase and decrease,
when $L$ increases. This is due to the onset of the  $\Sigma^0$ hyperon and
occurs for both  $\omega\rho$ and $\sigma\rho$ nonlinear
interactions, see Fig. \ref{figure:onset}. 
A similar
situation occurs for $U^N_\Xi=0$ and $-18$ MeV, however in this case 
$\Sigma^0$ sets in only for the $\omega\rho$ mixture and for
$L\lesssim 80$ MeV. The radii and strangeness fraction in
this case come close to the ones obtained with $U^N_\Xi=+18$ MeV. It
is interesting to notice that this is a
quite different behavior from the trend obtained with the $\sigma\rho$ nonlinear
term where the radius difference between $L=50$ and 110 MeV is
not larger than 1 km and the strangeness fraction for $L=50$ MeV is
half the one obtained with the $\omega\rho$ term, 
see plot (h) of the bottom panel of Fig.  \ref{figure:max_mass} and plot (d) of Fig. \ref{figure:max_radius}.  

\begin{figure}[htb]
% figura 10
\includegraphics[width=1.0\linewidth,angle=0]{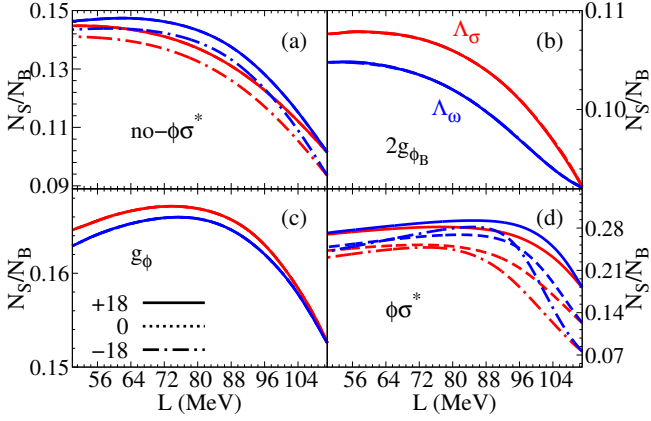}
\caption{
\label{figure:max_star_YS_4_param_-30}
(Colour online) Maximum mass star strangeness fraction versus $L$ for TM1-2 with $U_\Sigma^N=-30$ MeV and
all the other optical potentials (+18,0,-18) for $U_\Xi^N$ in MeV.
}
\end{figure}

In Fig. \ref{figure:max_star_YS_4_param_-30} the strangeness fraction
is plotted for $U_\Sigma^N=-30$ MeV. In this case the $\Sigma^-$
hyperon is the first to set in. Since a smaller $L$ favors the
formation of negatively charged hyperons, the general trend is a
decrease of the strangeness when $L$ increases, even for a
repulsive $YY$ interaction. For the attractive YY parametrization the
fast decrease of the strangeness for $L\gtrsim 90$ MeV is due to the
disappearance of the $\Sigma^0$ hyperon.

\begin{figure}[htb]
% figura 11
\includegraphics[width=1.0\linewidth,angle=0]{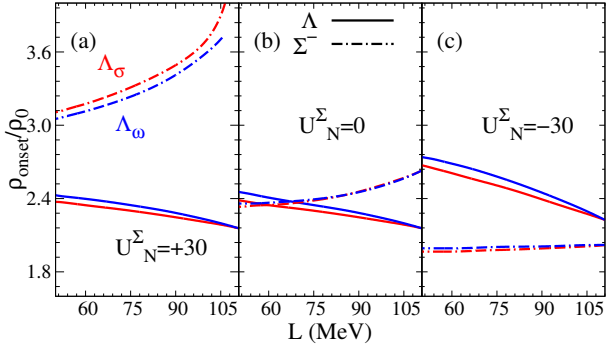}
\caption{
\label{figure:particle_onset_change_sigma}
(Colour online) Competition to first hyperon onset
between the $\Sigma^-$ and the $\Lambda$ elements of the lightest baryon octet when
we change $U^N_\Sigma$.}
\end{figure}

Fig. \ref{figure:particle_onset_change_sigma} clarifies
the existing competition between
the $\Lambda$ and the $\Sigma^-$ onset. 
An attractive $U_\Sigma^N$
potential favors the  $\Sigma^-$ onset. On the contrary a repulsive
$U_\Sigma^N$ favors the $\Lambda$, but for $U_\Sigma^N=0$ or close the
$L$ defines the hyperon onset: small $L$ favors the $\Sigma^-$. This
behavior was already discussed in \cite{providencia13}.
The other parametrizations of the hyperon-meson couplings give rise to
a similar behavior and, therefore, are not represented.

\begin{figure}[htb]
% figura 12
\includegraphics[width=1.0\linewidth,angle=0]{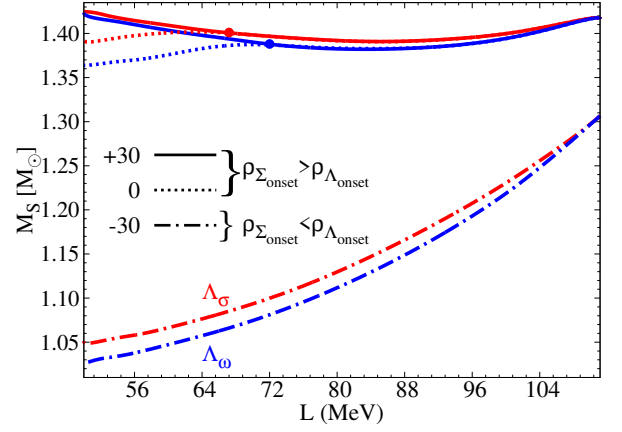}
\caption{
\label{figure:mass_of_strangeness_onset}
(Colour online) Star mass of onset of strangeness versus $L$ for TM1-2
for $U_\Xi^N=-18$ MeV and several values for  $U_\Sigma^N$.
For $U_\Sigma^N=0$, the full dots indicate the critical slope, $L=72$
MeV for $\Lambda_\omega$ and $L=67.2$ MeV for
$\Lambda_\sigma$, above which
$\rho_{\Sigma_\textrm{onset}}>\rho_{\Lambda_\textrm{onset}}$. For
$U_\Sigma^N=-30$ (+30) the first hyperon to appear is the $\Sigma^-$.
}
\end{figure}
The star mass of strangeness onset, \textit{i.e.}, the mass of the star
with a droplet of strangeness at the center and an infinitesimal
$N_S/N_B$, is plotted in
Fig. \ref{figure:mass_of_strangeness_onset}.

These curves do not
depend on the inclusion of strange mesons, which only act in  matter
with a finite amount of hyperons. The $\Lambda$ and $\Sigma^-$
hyperons compete to appear first
as $U^N_\Sigma$ and $L$ change. The repulsive $U^N_\Sigma=+30$ MeV or
even the $U^N_\Sigma=0$ MeV potential
favor the onset of the $\Lambda$ hyperon, as already referred above. However, in the last case,
for a low enough $L$,  the $\Sigma^-$ sets in first if $L\le 67.2$ MeV for $\sigma\rho$ or
$L\le 72$ MeV for the $\omega\rho$ non-linear term. Below those
critical values of $L$  the strangeness onset mass decreases with the
decrease of $L$. If  $U^N_\Sigma=-30$ MeV
the onset of the $\Sigma^-$ occurs always at a smaller density than
$\Lambda$. If the first hyperon to set in is the $\Lambda$ the  star
mass of strangeness onset is $\sim 1.4\,M_\odot$ and the dependence on
the slope $L$ is small. However, if  $U^N_\Sigma$ is attractive,
$\Sigma^-$ may be the first hyperon to set in first,  and, in this
case, the
strangeness onset star mass is sensitive to $L$.  For
$U^N_\Sigma=-30$ MeV this mass decreases from $\sim 1.3$ to  $1.05\,M_\odot$
when $L$ decreases from 111 to 50 MeV. We conclude that if the
$U_\Sigma^N$ potential is repulsive as experiments seem to indicate \cite{avraham_gal_hyperon_optical_potentials_2010}
we may expect strangeness in stars that {have} a mass at least as large as
$\sim 1.4 M_\odot$. Less massive stars  are totally determined by the
nucleonic properties of the EOS.

\begin{center}
\begin{figure}[htb]
% figura 13
\includegraphics[width=1\linewidth,angle=0]{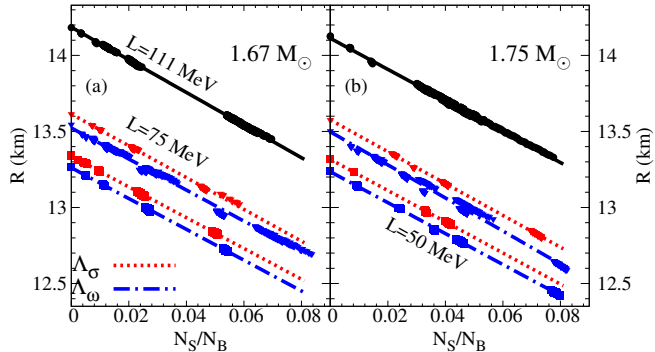}
\caption{
\label{figure:radii_vs_strangeness_all}
(Colour online) Radii versus total strangeness content of the stars
with masses $1.67M_\odot$ and $1.75M_\odot$
for  $L=111$ MeV (circles, full line), $L=75$ MeV (triangles),
$L=50$ MeV (squares). The straight lines are fitted to the data obtained.
The red symbols and red dotted line refers to the EOS with the $\sigma\rho$
term and the blue symbols and dashed-dotted to the EOS with the
$\omega\rho$ term.  The points represented include information from
all the hyperon-meson couplings considered.
}
\end{figure}
\end{center}

\begin{table}[htb]
\caption{
\label{table:table_of_all_strangeness_straight_lines}
Parameters (slope $\eta_S= \frac{dR}{d(N_S/N_B)}$ and zero strangeness
radius $R_0=R(Y_S=0)$) for the straight lines that describe the star radius versus the
total strangeness fraction for four values of the star mass: 1.44,
1.60, 1.65 and 1.75 $M_\odot$,  see Fig. \ref{figure:radii_vs_strangeness_all}. The uncertainties in the fits
are below $6\%$ in the slope and below $0.02\%$ in $R_0$.
}
\begin{tabular}{lcccccccc}
\hline 
 &   \multicolumn{2}{ c }{1.44 $M_\odot$}  & \multicolumn{2}{c }{1.60 $M_\odot$}  & \multicolumn{2}{c }{1.67 $M_\odot$}  & \multicolumn{2}{c }{1.75 $M_\odot$}  \\
$L$           & $\eta_S$        & $R_0$& $\eta_S$        & $R_0$& $\eta_S$        & $R_0$& $\eta_S$        & $R_0$\\
(MeV)                                                            &(km)&     (km)  &(km)&     (km)  &(km)&     (km)  &(km)&     (km)  \\
\hline
  111                                     &       10.30                   &      14.32       &       10.69                   &      14.23    &       10.72&      14.19  &       10.22                   &      14.11\\

  75 ($\Lambda_\sigma$)   &       10.82                   &      13.66  &       10.09                  &      13.63   &       10.46                   &      13.61  &       10.47                   &      13.58 \\
 75 ($\Lambda_\omega$)  &       10.84                   &
             13.59 &       10.20                   &      13.55  &
             10.21                   &      13.52  &       10.87                 &      13.50\\

50 ($\Lambda_\sigma$)   &       10.22                   &      13.32  &       10.16                &      13.34 &       10.12                  &      13.34  &       10.26                   &      13.32\\
 50  ($\Lambda_\omega$)  &       10.33                   &      13.27 &       10.31                   &      13.28  &       10.13                   &      13.26  &       10.17                  &      13.24\\
\hline
\end{tabular}
\end{table}

We have seen that strangeness is present in stars with masses (above or equal) to
$\approx 1M_\odot - 1.42M_\odot$ depending on $U^N_\Sigma$.
$L$ strongly influences neutron star radius
and the strangeness content. To understand the effect
of strangeness and $L$ on the star  radius, we have fixed the star
mass and calculated the radius changing 
 both $L$ and strangeness. In
 Fig. \ref{figure:radii_vs_strangeness_all} we plot the results
 obtained for a 1.67  $M_\odot$ and a 1.75 $M_\odot$ star.
It is clearly seen that the  radius decreases linearly with the
increase of the
strangeness fraction for all the nucleonic EOS shown, which differ in
the value of $L$ and non-linear $\rho$ term in the Lagrangian.
We have fitted the data for the star radius as a function of the
strangeness fraction through
\begin{equation}
R= \eta_S\, Y_s +R_0  (\mbox{km}),
\label{R}
\end{equation}
where $R_0=R(Y_S=0)$ and $\eta_S=\frac{dR}{d Y_S}$, $Y_S=N_S/N_B$.
The parameters of the fitted lines for four values of the
star mass are listed in Table
\ref{table:table_of_all_strangeness_straight_lines}.
The values of the slope are approximately equal within the uncertainty
of the slope and agree with the values obtained previously in \cite{providencia13}.

Radius estimates inferred from photospheric radius expansion bursts and thermal emissions from
quiescent low-mass X-ray binaries and isolated neutron stars using Bayesian techniques \cite{steiner2014} indicate that the radius of a
1.4$M_\odot$ star is 12.1$\pm 1.1$ km.  If 1.4$M_\odot$ stars are
purely nucleonic the radius gives  information on the
density dependence of the symmetry energy, and small values of $L$ are
favored. According to our study if the $U_\Sigma^N$ is repulsive, it
is very likely that 1.4$M_\odot$ stars are nucleonic stars. However, if $U_\Sigma^N$ is attractive, small
values of $L$ favor an early appearance of strangeness and the radius
is determined by both the density dependence of the symmetry energy
and the strangeness content.

\section{CONCLUSIONS}
\label{section:conclusions}

In the present study we  discuss the joint effect of the density
dependence of the symmetry energy and hyperon interaction on several
properties of neutron stars, including radius, gravitational mass,
strangeness content and star mass at strangeness onset. This was carried out within the
relativistic mean field framework. The density dependence of the
symmetry energy was modelized 
 including a non-linear $\omega\rho$ or
$\sigma\rho$ term in the Lagrangian density. We have changed the symmetry
energy slope at saturation between 50 and 110 MeV, taking care that
the parametrizations obtained do no predict neutron star matter
that saturates at a finite density.  

For the hyperon meson interaction
we have considered the isoscalar vector mesons $\omega$,$\phi$-coupling constants defined
by the SU(6) symmetry,
while the $\rho$-meson coupling is defined by the  isospin
symmetry of the hyperon, taking the nucleon coupling as reference. In
order to get a  quite repulsive $YY$ interaction, we have also considered a parametrization with
broken SU(6) symmetry, and took a parametrization with $g_\phi> g_\phi(SU(6))$.
To define the hyperon-$\sigma$ meson couplings
we use the hyperonic potential in symmetric  nuclear matter at saturation
and take $U^N_\Lambda=-28$ MeV. Since $U^N_\Sigma$
and  $U^N_\Xi$ are not  well constrained we have considered several
possibilities from attractive to  repulsive potentials for these two
hyperonic potentials. The hyperon-$\sigma^*$ coupling was chosen
non-zero in a single parametrization which gives rise to a quite attractive $YY$
interaction.  

Next we summarize some of the main conclusions.

Nucleonic stars have smaller radii for EOS with smaller values of
$L$ as shown in \cite{carriere2003}, and this effect is larger if the
non-linear $\omega\rho$ instead of the $\sigma\rho$ term is used.
These two terms implement different density dependencies on the
symmetry energy due to the different behavior of the $\sigma$ and
$\omega$ mesons, the first one saturates at high densities while the
second one increases almost linearly with density. Consequently  the
$\sigma\rho$ ($\omega\rho$) term has a stronger effect below (above)
saturation density.
 The
gravitational mass is not affected by more than $\sim 3\%$ when $L$ changes
from $50$ to $110$ MeV. 

Including hyperons will soften the EOS, however quantitatively the
effect depends a lot on the hyperon parametrization chosen. We have
shown that $2 M_\odot$ are obtained if the $YY$ interaction is
repulsive enough, as shown in previous works
\cite{haensel11,schaffner12b,sedrakian12,oertel2014,debora14}. 
Within the TM1 parametrization chosen, it could be shown that if the
hyperonic potential $U_\Sigma^N$ is repulsive there will only exist
strangeness in the core for stars with $M\gtrsim 1.4 M_\odot$, and
that the first hyperon to set in is the $\Lambda$. On the
other hand an attractive  $U_\Sigma^N$ favors the onset of the
$\Sigma^-$ before the $\Lambda$ and  a $1
M_\odot$ star may already have strangeness in its core. If the
potential  $U_\Sigma^N$ is close to zero the value of $L$ will define
whether the $\Lambda$ or the $\Sigma^-$ will first set in, with a
small $L$ favoring the appearance of the $\Sigma^-$.

We could confirm the results of \cite{providencia13} that star radius
depends linearly on the total strangeness fraction, and we have shown
that the slope of this linear correlation does not depend neither in
$L$ nor in the star
mass and is $\sim 10.5$ km/$\eta_S$. 

The existence of strangeness in the core of a 1.4 $M_\odot$ star may
have to be  considered when the $\Sigma$ potential in nuclear matter
is not repulsive. In this case, the astrophysical observations  will
give information not only on the slope $L$ of the symmetry energy but
also on the strangeness content that may be estimated using
expression (\ref{R}).

We have also shown that while in nucleonic stars reducing the slope
$L$ from 110 to 50 MeV may give rise to a decrease of the maximum mass
star radius of $\sim 0.5$ km, for hyperonic stars this reduction can
be of the order of 1 km or even larger if an attractive $YY$ is considered.

Finally it should also be referred that to understand the structure of
a neutron star it is not enough to constrain properties at saturation
density, but information on the  density dependence of the equation of
state, in particular, the symmetry energy is also needed. Taking two
different parametrizations for the isospin channel of the equation of
state it was shown  the density dependence of the symmetry at high
density influences properties such as the mass, radius or strangeness
content of the star.

In this study we were very conservative and we have only considered
for the hyperon couplings one possibility that did not satisfy the
SU(6) symmetry. In \cite{oertel2014} several parametrizations have
been proposed that do not satisfy either the SU(6) symmetry and even
with a YY attractive interaction in some of the channels it was
possible to describe a 2 $M_\odot$ star. With the present work we just
want to show in a more systematic way that it is important to put
constraints on the hyperon interactions before the interpretation of
observational data may be used to set constraints on the 
equation of state.

\acknowledgments

This work is partly supported by the project PEst-OE/FIS/UI0405/2014 developed under the initiative QREN financed by the UE/FEDER through the program
COMPETE-"Programa Operacional Factores de Competitividade", and by "NewCompstar", COST Action MP1304.

\section*{Appendix}
 In this section we display the properties of maximum mass stars
 calculatd with the EOS presented.

\begin{table*}[htb]
\caption{
\label{table:hyp1}
Maximum star properties for the TM1-2 model. The hyperon potentials are
$U_\Sigma^N=+30$ MeV and $U^N_\Xi=+18,\, 0, \, -18$ MeV. The minimum gravitational mass star is also listed
in the intermediate critical $L_c$. $M_S$ is the star mass of onset of strangeness for a certain $L$.
}
\begin{tabular}{cccccccccccccccccccccccccccc}
\hline
&\multicolumn{7}{c}{$U^N_\Xi=+18$}&$\phantom{c}$&\multicolumn{7}{c}{$U^N_\Xi=0$}&$\phantom{c}$&\multicolumn{7}{c}{$U^N_\Xi=-18$} \\
&   $L$      & $M_g$  &   $M_b$    &   $R$      &$\varepsilon_0$&
  $\eta_S$ &$\rho_c$& $M_S$  &  $L$    & $M_g$  &   $M_b$    &   $R$      &$\varepsilon_0$&
  $\eta_S$ &$\rho_c$& $M_S$  & $L$    & $M_g$  &   $M_b$    &   $R$      &$\varepsilon_0$&
  $\eta_S$ &$\rho_c$& $M_S$  & \\
\hline
&\multicolumn{22}{c}
{$no$-$\phi\sigma^*$}\\
& 111      & 1.80   &    2.02    &   13.19    &   4.30      &   0.087   & 0.75 & 1.41  &     & 1.77   &    1.99    &   13.34    &   4.07      & 0.077    & 0.72  & 1.41 &      & 1.72   &    1.92    &   13.43    &   3.92      &  0.076   & 0.69  & 1.41\\
$\Lambda_\omega$ &  96       & 1.79   &    2.01    &   12.88    &   4.51      &  0.088    & 0.78 & 1.38 &    & 1.75   &    1.97    &   12.99    &   4.34      &  0.083   &  0.76 & 1.37 &     & 1.70   &    1.91    &   13.00    &   4.28      &  0.091   & 0.75  & 1.38\\
   &  50       & 1.81   &    2.07    &   12.47    &   4.67      &   0.079   & 0.81 & 1.41 &      & 1.79   &    2.03    &   12.56    &   4.49      &  0.077   & 0.78  & 1.41 &      & 1.74   &    1.96    &   12.50    &   4.535      &  0.093   & 0.79  & 1.41\\
$\Lambda_\sigma$ &  96       & 1.79   &    2.02    &   12.95    &   4.45      &   0.088   & 0.77 & 1.38 &       & 1.76   &    1.98    &   13.06    &   4.26      &  0.082   & 0.75 & 1.38 &      & 1.71   &    1.92    &   13.09    &   4.18      &  0.088   & 0.74 & 1.38\\
   &  50       & 1.82   &    2.07    &   12.55    &   4.57      &   0.079   & 0.79 & 1.41&      & 1.79   &    2.03    &   12.64    &   4.40      &  0.076   & 0.77 & 1.41  &      & 1.74   &    1.97    &   12.60    &   4.41      &  0.090   & 0.77 & 1.41\\

&\multicolumn{22}{c}
{$2g_\phi$}\\
                 &  111      & 2.12   &    2.45    &   12.31    &   5.48      &   0.078   & 0.87 & 1.39 &   & 2.12   &    2.45    &   12.30    &   5.48      & 0.078     & 0.87 & 1.42 &    & 2.12   &    2.45    &   12.30    & 5.48        & 0.078     & 0.87 & 1.39\\
$\Lambda_\omega$ &  79       & 2.08   &    2.42    &   12.01    &   5.63      &   0.067   & 0.90 & 1.38 &      & 2.08   &    2.42    &   12.01    &   5.63      &   0.067   & 0.90 & 1.38 &      & 2.08   &    2.42    &   12.01    & 5.63        & 0.067     & 0.90 & 1.38\\
               &  50       & 2.09   &    2.44    &   11.91    &   5.60      &   0.062   & 0.90 & 1.41 &      & 2.09   &    2.44    &   11.91    &   5.60      &   0.062   & 0.90 & 1.41 &    & 2.09   &    2.44    &   11.92    & 5.60      & 0.062     & 0.90 & 1.41\\
$\Lambda_\sigma$ &  68       & 2.10   &    2.45    &    12.03   &   5.57     &   0.068   & 0.89 & 1.38 &     & 2.10   &    2.45    &    12.03   &   5.57     &   0.068   & 0.89 & 1.38 &    & 2.10   &    2.45    &    12.03   & 5.57       & 0.068& 0.89 & 1.38\\
                &  50       & 2.11   &    2.46    &    11.97   &   5.54      &   0.064   & 0.89 & 1.41 &    & 2.11   &    2.46    &    11.97   &   5.54      &   0.064   & 0.89 & 1.41 &    & 2.11   &    2.49    &    11.97   & 5.54       & 0.065     & 0.89 & 1.41\\

 &\multicolumn{22}{c}
{$g_\phi$}\\
             &  111      & 1.80   &    2.02    &   13.19    &   4.30
             &   0.087  & 0.75 & 1.41 &  & 1.96   &    2.23    &
             12.44    &   5.31      &   0.125   & 0.87 & 1.41 & & 1.94   &    2.21    &   12.41    &   5.36      &  0.138    & 0.88 & 1.41\\
 $\Lambda_\omega$ &  86       & 1.94   &    2.22    &   12.14    &   5.50       &   0.111  & 0.90 & 1.37 &      & 1.94   &    2.22    &   12.14    &   5.48      &  0.113    & 0.90 & 1.37 &  90       & 1.92   &    2.19    &   12.10    &   5.58      &  0.133    & 0.92 & 1.36\\
              &  50       & 1.81   &    2.07    &   12.47    &   4.67
              &   0.079  & 0.81 & 1.41 &        & 1.95   &    2.25
              &   11.99    &   5.44      &  0.100    & 0.90 & 1.41 &  & 1.94   &    2.23    &   11.94    &   5.53      &  0.118    & 0.91 & 1.40 \\
$\Lambda_\sigma$ &  80       & 1.96 &    2.24    &   12.18    &
5.44      &   0.112  & 0.90 & 1.38 &  89       & 1.95   &    2.23
&   12.23    &   5.41     &   0.119   & 0.89 & 1.38 &  92       & 1.93
&    2.21    &   12.18    &   5.51      &  0.137    & 0.91 & 1.38\\
    &  50       & 1.82   &    2.07    &   12.55    &   4.57       &
    0.079  & 0.79 & 1.42 &      & 1.96   &    2.26    &   12.06    &
    5.38      &   0.103   & 0.89 & 1.41  &      & 1.95   &    2.24    &   11.99    &   5.47      &  0.121    & 0.90 & 1.41\\

&\multicolumn{22}{c}
{$\phi\sigma^*$}\\
           &  111      & 1.80   &    2.02    &   13.19    &   4.30       &   0.087   & 0.75& 1.41 &    & 1.61   &    1.78    &   13.85    &   3.30      &   0.038   & 0.60 & 1.39 &  & 1.58   &    1.74    &   13.94    &   3.15      &   0.032   & 0.58 & 1.39\\
$\Lambda_\omega$ &  89       & 1.60   &    1.78    &   12.78    &
4.40       &   0.108   & 0.78& 1.38 &       & 1.58   &    1.76
&   13.27    &   3.66     &   0.052   & 0.66 & 1.37  & & 1.55   &    1.72    &   13.32    &   3.58     &   0.053   & 0.65 & 1.35\\
                &  50       & 1.81   &    2.07    &   12.47    &
                4.67       &   0.079   & 0.81& 1.41&      & 1.61   &
                1.80    &   12.31    &   4.68      &   0.119   & 0.83
                & 1.41 &   & 1.59   &    1.77    &   12.28    &   4.73      &   0.130   & 0.83 & 1.40\\
$\Lambda_\sigma$ &  89      & 1.61   &    1.79    &   12.98    &
4.10       &   0.090   & 0.73& 1.39 &       & 1.59   &    1.77    &   13.39    &   3.56      &   0.048   & 0.65 & 1.38 & &       1.56 &        1.73 &        13.46 &        3.45 &        0.047&  0.63 & 1.36\\
               &  50       & 1.82   &    2.07    &   12.55    &
               4.57       &   0.079   & 0.79& 1.39 &      & 1.62   &    1.81    &   12.86    &   3.82      &   0.058   & 0.69 & 1.39  &    & 1.59 &       1.78 &        12.84 &        3.81 &         0.064 &  0.69 & 1.41\\
\hline
\end{tabular}
\end{table*}

\begin{table*}[htb]
\caption{
\label{table:hyp2}
Maximum star properties for the TM1-2 model. The hyperon potentials are
$U_\Sigma^N=0$ MeV and $U^N_\Xi=+18,\, 0,\, -18$ MeV. The minimum gravitational mass star is also listed
in the intermediate critical $L_c$. $M_S$ is the star mass of onset of strangeness for a certain $L$.
}
\begin{tabular}{cccccccccccccccccccccccccccc}
\hline
&\multicolumn{7}{c}{$U^N_\Xi=+18$}&$\phantom{c}$&\multicolumn{7}{c}{$U^N_\Xi=0$}&$\phantom{c}$&\multicolumn{7}{c}{$U^N_\Xi=-18$} \\
&   $L$      & $M_g$  &   $M_b$    &   $R$      &$\varepsilon_0$&
  $\eta_S$ &$\rho_c$& $M_S$  &  $L$    & $M_g$  &   $M_b$    &   $R$      &$\varepsilon_0$&
  $\eta_S$ &$\rho_c$& $M_S$  & $L$    & $M_g$  &   $M_b$    &   $R$      &$\varepsilon_0$&
  $\eta_S$ &$\rho_c$& $M_S$  & \\
\hline
&\multicolumn{22}{c}
{$no$-$\phi\sigma^*$}\\
&  111      & 1.78   &    2.00    &   13.06    &   4.44      &  0.098   & 0.77 & 1.41  &  & 1.76   &    1.98    &   13.27    &   4.15      &  0.081   & 0.73 & 1.41 &    & 1.72   &    1.92    &   13.42    &   3.93      &  0.076   & 0.70 & 1.41\\
$\Lambda_\omega$ &  86       & 1.74   &    1.96    &   12.43    &   5.02      &  0.118   & 0.86 & 1.37&    & 1.73   &    1.95    &   12.60    &   4.75     &  0.103   & 0.82 & 1.37 &  90       & 1.70   &    1.90    &   12.82    &   4.48      &  0.099   & 0.78 & 1.37\\
&  50       & 1.76   &    1.99    &   12.09    &   5.19      &  0.119   & 0.89 & 1.35 &      & 1.75   &    1.99    &   12.22    &   4.96      &  0.107   & 0.86 & 1.35 &    & 1.73   &    1.95    &   12.33    &   4.78      &  0.103   & 0.83 & 1.35\\
$\Lambda_\sigma$ &  83       & 1.75   &    1.98    &   12.51    &   4.89      &  0.115   & 0.84 & 1.37&  86       & 1.74   &    1.96    &   12.74    &   4.59    &  0.099   & 0.80 & 1.38 &  92       & 1.71   &    1.91    &   12.99    &   4.28      &  0.092   & 0.75 & 1.38\\
&  50       & 1.77   &    2.00    &   12.19    &   5.05      &  0.116   & 0.87 & 1.38 &  & 1.76   &    1.99    &   12.33    &   4.841      &  0.104   & 0.84 & 1.38&   & 1.73   &    1.96    &   12.46    &   4.62      &  0.099   & 0.81 & 1.38\\

&\multicolumn{22}{c}
{$2 g_\phi$}\\              

&  111      & 2.11   &    2.44    &   12.31    & 5.47       & 0.080    & 0.87 & 1.41 &  & 2.11   &    2.44    &   12.31    & 5.47        & 0.080    & 0.87 & 1.41 &     & 2.11   &    2.44    &   12.31    & 5.47        & 0.080    & 0.87 & 1.41\\
$\Lambda_\omega$ &  73       & 2.05   &    2.38    &   11.88    & 5.78& 0.076    & 0.93 & 1.37&      & 2.05   &    2.38    &   11.88    & 5.78      & 0.076    & 0.93 & 1.38 &        & 2.05   &    2.38    &   11.89    & 5.78        & 0.076    & 0.93 & 1.38\\
 &  50       & 2.06   &    2.40    &   11.79    & 5.77        & 0.074    & 0.93 & 1.35 &      & 2.06   &    2.40    &   11.79    & 5.77        & 0.074    & 0.93 & 1.35 &       & 2.06   &    2.40    &   11.79    & 5.77        & 0.074    & 0.93 & 1.35\\
$\Lambda_\sigma$ &  50       & 2.08   &    2.42    &   11.86    & 5.69& 0.076    & 0.92 & 1.38 &      & 2.08   &    2.42    &   11.86    & 5.69        & 0.076    & 0.92 & 1.38&        & 2.08   &    2.42    &   11.86    & 5.69        & 0.076    & 0.92 & 1.38\\

&\multicolumn{22}{c}
{$g_\phi$}\\
  &  111      & 1.94   &    2.21    &   12.40    & 5.37       & 0.129    & 0.88 & 1.41 &   & 1.94   &    2.21    &   12.40    & 5.37       & 0.128    & 0.88 & 1.41 &     & 1.93   &    2.20    &   12.42    & 5.35        & 0.134    & 0.88 & 1.40\\
$\Lambda_\omega$ &  79       & 1.90   &    2.17    &   11.92    & 5.76        & 0.127    & 0.95 & 1.37 &      & 1.90   &    2.17    &   11.92    & 5.76        & 0.127    & 0.95 & 1.37 &  84       & 1.90   &    2.16    &   11.96    & 5.73        & 0.129    & 0.94 & 1.37\\
&  50       & 1.91   &    2.19    &   11.78    & 5.77        & 0.121    & 0.95 & 1.33 &     & 1.91   &    2.19    &   11.77    & 5.77        & 0.121    & 0.95 & 1.33 &      & 1.91   &    2.19    &   11.78    & 5.76        & 0.121    & 0.95 & 1.33\\
$\Lambda_\sigma$ &  71       & 1.91   &    2.19    &   11.97    & 5.66        & 0.127    & 0.93 & 1.39 &       & 1.91   &    2.19    &   11.96    & 5.66        & 0.127    & 0.93 & 1.39 &  75       & 1.91   &    2.19    &   12.01    & 5.64        & 0.130    & 0.93 & 1.38\\
     &  50       & 1.92   &    2.20    &   11.85    & 5.67       & 0.123    & 0.93 & 1.36 &     & 1.92   &    2.20    &   11.85    & 5.67        & 0.123    & 0.93 & 1.3&      & 1.92   &    2.20    &   11.85    & 5.66        & 0.123    & 0.93 & 1.36\\

&\multicolumn{22}{c}
{$\phi\sigma^*$}\\
&  111      & 1.59   &    1.76    &   13.44    & 3.78        & 0.079    & 0.68 & 1.39  &   & 1.59   &    1.76    &   13.79    & 3.33        & 0.044    & 0.61 & 1.40 &       & 1.57   &    1.73    &   13.98    & 3.09        & 0.027    & 0.57 & 1.39\\
$\Lambda_\omega$ &  91       & 1.56   &    1.73    &   12.39    &5.03        & 0.154    & 0.88 & 1.35 &      & 1.55   &    1.72    &   12.61    & 4.68        & 0.132    & 0.83 & 1.37&       & 1.53   &    1.69    &   12.81    & 4.34        & 0.115    & 0.77 & 1.35\\
 &  50       & 1.59   &    1.78    &   11.83    & 5.54        & 0.171    & 0.96 & 1.34 &          & 1.58 &        1.77 &        12.00 &      5.24 &    0.152 &   0.91 & 1.33&       & 1.56   &    1.74    &   12.11    & 5.02      & 0.144    & 0.88 & 1.33\\
$\Lambda_\sigma$ &  89       & 1.57   &    1.74    &   12.55    & 4.79        & 0.141    & 0.84 & 1.38 &     & 1.56   &    1.73    &   12.78    & 4.42       & 0.118    & 0.79 & 1.38&       & 1.54   &    1.70    &   13.03    & 4.02        & 0.094    & 0.72 & 1.38\\
 &  50       & 1.59   &    1.78    &   11.95    & 5.32        & 0.163
 & 0.92 & 1.36 &      & 1.59   &    1.78    &   11.95    & 5.32
 & 0.163    & 0.92 & 1.36&       & 1.57   &    1.75    &   12.24    &
 4.81        & 0.135    & 0.85 & 1.35\\
\hline
\end{tabular}

\end{table*}

\begin{table*}[htb]
\caption{
\label{table:hyp3}
Maximum star properties for the TM1-2 model. The hyperon potentials are
$U_\Sigma^N=-30$ MeV and $U^N_\Xi=+18,\, 0,\, -18$ MeV. The minimum gravitational mass star is also listed
in the intermediate critical $L_c$. $M_S$ is the star mass of onset of strangeness for a certain $L$.
}
\begin{tabular}{cccccccccccccccccccccccccccc}
\hline
&\multicolumn{7}{c}{$U^N_\Xi=+18$}&$\phantom{c}$&\multicolumn{7}{c}{$U^N_\Xi=0$}&$\phantom{c}$&\multicolumn{7}{c}{$U^N_\Xi=-18$} \\
&   $L$      & $M_g$  &   $M_b$    &   $R$      &$\varepsilon_0$&
  $\eta_S$ &$\rho_c$& $M_S$  &  $L$    & $M_g$  &   $M_b$    &   $R$      &$\varepsilon_0$&
  $\eta_S$ &$\rho_c$& $M_S$  & $L$    & $M_g$  &   $M_b$    &   $R$      &$\varepsilon_0$&
  $\eta_S$ &$\rho_c$& $M_S$  & \\
\hline
&\multicolumn{22}{c}
{$no$-$\phi\sigma^*$}\\
 &  111      & 1.69   &    1.89    &   13.11    &   4.37      &  0.101   & 0.77 & 1.29 &   & 1.69   &    1.89    &   13.11    &   4.37      &  0.101   & 0.77 & 1.29 &   & 1.68   &    1.87    &   13.21    &   4.23      &  0.093     & 0.74 & 1.30\\
 $\Lambda_\omega$ &  86       & 1.65   &    1.84    &   12.28    &   5.22      &  0.140   & 0.90 & 1.13 &     & 1.65   &    1.84    &   12.28    &   5.22      &  0.140   & 0.90 & 1.13 &  89       & 1.64   &    1.83    &   12.41    &   5.07      &  0.133     & 0.88 & 1.14\\
&  50       & 1.67   &    1.88    &   11.88    &   5.49      &  0.146   & 0.94 & 1.02 &     & 1.67   &    1.88    &   11.87    &   5.49      &  0.146   & 0.94 & 1.02 & & 1.67   &    1.88    &   11.91    &   5.47      &  0.144     & 0.94 & 0.99\\
$\Lambda_\sigma$ &  83       & 1.66   &    1.86    &   12.38    &   5.06      &  0.136   & 0.87 & 1.13 &     & 1.66   &    1.86    &   12.39    &   5.06      &  0.136   & 0.87 & 1.13 &  & 1.65   &    1.85    &   12.49    &   4.93      &  0.129     & 0.85 & 1.14\\
&  50       & 1.67   &    1.89    &   11.97    &   5.37      &  0.145   & 0.92 & 1.04 &     & 1.67   &    1.89    &   11.97    &   5.37      &  0.145   & 0.92 & 1.0&    & 1.67   &    1.88    &   12.01    &   5.30      &  0.141     & 0.91 & 1.02\\

&\multicolumn{22}{c}
{$2 g_\phi$}\\              
 &  111      & 2.09   &    2.41    &   12.16    & 5.65        & 0.096    & 0.90 & 1.30 &  & 2.09   &    2.41    &   12.16    & 5.65        & 0.096    & 0.90 & 1.28 &  & 2.09   &    2.41    &   12.16    &   5.65      &  0.096     & 0.90 & 1.30\\
$\Lambda_\omega$ &  70       & 2.02   &    2.33    &   11.61    & 6.13        & 0.102    & 0.98 & 1.05 &    & 2.02   &    2.34    &   11.59    & 6.14        & 0.102    & 0.98 & 1.04 &    & 2.02   &    2.33    &   11.60    &   6.13      &  0.102     & 0.98 & 1.05\\
&  50       & 2.02   &    2.35    &   11.51    & 6.15        & 0.102    & 0.98 & 0.99 &     & 2.02   &    2.35    &   11.51    & 6.15        & 0.102    & 0.98 & 0.99 &    & 2.02   &    2.35    &   11.51    &   6.15      &  0.102     & 0.98 & 0.99\\
$\Lambda_\sigma$ &  50       & 2.04   &   2.37     &   11.59    &
6.05        & 0.104    & 0.97 & 1.01 &     & 2.04   &   2.37     &   11.59    & 6.05        & 0.104    & 0.97 & 1.01 &    & 2.04   &    2.37    &   11.58    &   6.05      &  0.104     & 0.97 & 1.01\\

&\multicolumn{22}{c}
{$ g_\phi$}\\   
&  111      & 1.89   &    2.15    &   12.24    & 5.584        & 0.152    & 0.92 & 1.30 &     & 1.89   &    2.15    &   12.24    & 5.58        & 0.152    & 0.92 & 1.30 &   & 1.89   &    2.15    &   12.24    &   5.58      &  0.152     & 0.92 & 1.30\\
$\Lambda_\omega$ &  84       & 1.84   &    2.09    &   11.60    & 6.252        & 0.166    & 1.02 & 1.11 &     & 1.84   &    2.09    &   11.60    & 6.25        & 0.166    & 1.02 & 1.11 &  & 1.84   &    2.09    &   11.60    &   6.25      &  0.166     & 1.02 & 1.11\\
&  50       & 1.85   &    2.12    &   11.38    & 6.332        & 0.163    & 1.03 & 1.01 &     & 1.85   &    2.12    &   11.39    & 6.33        & 0.163    & 1.03 & 1.01 &   & 1.85   &    2.12    &   11.38    &   6.33      &  0.163     & 1.03 & 1.01\\
$\Lambda_\sigma$ &  73       & 1.86   &    2.12    &   11.63    & 6.150        & 0.167    & 1.00 & 1.08 &     & 1.86   &    2.12    &   11.62    & 6.15        & 0.167    & 1.00 & 1.08 &  & 1.86   &    2.12    &   11.63    &   6.15      &  0.167     & 1.00 & 1.08\\
 &  50       & 1.86   &    2.13    &   11.47    & 6.207        & 0.165    & 1.01 & 1.01&     & 1.86   &    2.13    &   11.48    & 6.21        & 0.165    & 1.01 & 1.04&   & 1.86   &    2.13    &   11.47    &   6.21      &  0.165     & 1.01 & 1.01\\

&\multicolumn{22}{c}
{$\phi\sigma^*$}\\
   &  111      & 1.57   &    1.74    &   12.53    & 5.035        & 0.180    & 0.88 & 1.27 &   & 1.56   &    1.73    &   13.06    & 4.27        & 0.122    & 0.76 & 1.27 &   & 1.54   &    1.70    &   13.52    &   3.65      &  0.075     & 0.66 & 1.29\\
$\Lambda_\omega$ &  102      & 1.56   &    1.73    &   11.65    & 6.375        & 0.261    & 1.07 & 1.21 &     & 1.55   &    1.71    &   12.02    & 5.70        & 0.219    & 0.98 & 1.19 &     & 1.52   &    1.68    &   12.40    &   5.07      &  0.181     & 0.89 & 1.19\\
&  50       & 1.60   &    1.79    &   10.95    & 7.100        & 0.272    & 1.18 & 0.99 &    & 1.59   &    1.78    &   11.17    & 6.64        & 0.249    & 1.11 & 1.00 &    & 1.57   &    1.76    &   11.30    &   6.38      &  0.241     & 1.08 & 1.00\\
$\Lambda_\sigma$ &  102      & 1.57   &    1.73    &   11.88    & 5.982        & 0.241    & 1.02 & 1.23 &     & 1.55   &    1.72    &   12.32    & 5.23        & 0.189    & 0.91 & 1.23 &  99       & 1.53   &    1.69    &   12.60    &   4.86      &  0.166     & 0.85 & 1.21\\
 &  50       & 1.60   &    1.79    &   11.04    & 6.930        & 0.270
 & 1.15 & 1.03 &      & 1.59   &    1.78    &   11.29    & 6.42
 & 0.243    & 1.08 & 1.04 &     & 1.57 &       1.75 &     11.45 &
 6.16 &      0.231 &       1.04 & 1.01\\
\hline
\end{tabular}
\end{table*}

\end{document}